\documentclass[prd,preprintnumbers,twocolumn,amsmath,nofootinbib,amssymb]{revtex4}

\usepackage{graphicx,color,dcolumn,booktabs,bm}
\usepackage{longtable,lscape}
\usepackage{makecell}
\usepackage{txfonts}
\usepackage{overpic}
\usepackage{amssymb}
\usepackage{epstopdf}
\usepackage{indentfirst}
\usepackage{feynmf}   
\usepackage{slashed}  
\usepackage{cases}
\usepackage{color}
\usepackage{float}
\usepackage{multirow}
\usepackage{ulem}
\usepackage{enumerate}
\usepackage{graphicx,color,dcolumn,booktabs,bm}
\usepackage{epsfig,dsfont,amssymb,amsmath,amsfonts,amsbsy,mathrsfs}

\graphicspath{{Figures/}} %

\usepackage{hyperref}
\hypersetup{colorlinks,citecolor=blue,anchorcolor=red,menucolor=red, linkcolor=red,filecolor=red,runcolor=red,urlcolor=blue,frenchlinks=true}

\makeatletter
\@addtoreset{equation}{section}
\makeatother

\allowdisplaybreaks

\begin{document}

\title{Predicting charmed-strange molecular tetraquarks with \\$K^{(*)}$ and $T$-doublet charmed or anticharmed meson}

\author{Fu-Lai Wang$^{1,2,3,4}$}
\email{wangfulai@lzu.edu.cn}
\author{Si-Qiang Luo$^{1,2,3,4}$}
\email{luosq15@lzu.edu.cn}
\author{Xiang Liu$^{1,2,3,4}$\footnote{Corresponding author}}
\email{xiangliu@lzu.edu.cn}
\affiliation{$^1$School of Physical Science and Technology, Lanzhou University, Lanzhou 730000, China\\
$^2$Lanzhou Center for Theoretical Physics, Key Laboratory of Theoretical Physics of Gansu Province, Key Laboratory of Quantum Theory and Applications of MoE, Gansu Provincial Research Center for Basic Disciplines of Quantum Physics, Lanzhou University,
Lanzhou 730000, China\\
$^3$MoE Frontiers Science Center for Rare Isotopes, Lanzhou University, Lanzhou 730000, China\\
$^4$Research Center for Hadron and CSR Physics, Lanzhou University and Institute of Modern Physics of CAS, Lanzhou 730000, China}

\begin{abstract}
In this work, we first present a systematic investigation of the $T_{\bar{c}\bar{s}}$-type charmed-strange molecular tetraquark candidates composed of a $K^{(*)}$ meson and a $T$-doublet anticharmed meson using the one-boson-exchange model, which exhibit exotic flavor content $\bar{c}\bar{s} q q$. Our results suggest that the $K^* \bar D_1$ states with $I(J^P)=0(0^-,\,1^-)$ and the $K^* \bar D_2^*$ states with $I(J^P)=0(1^-,\,2^-)$ represent the most promising candidates of the $T_{\bar{c}\bar{s}}$-type charmed-strange molecular tetraquarks, while the coupled $K \bar D_1 / K^* \bar D_1 / K^* \bar D_2^*$ system with $I(J^P)=0(1^-)$ and the coupled $K \bar D_2^* / K^* \bar D_1 / K^* \bar D_2^*$ system with $I(J^P)=0(2^-)$ can only be regarded as the possible candidates of the $T_{\bar{c}\bar{s}}$-type charmed-strange molecular tetraquarks. We further extend our analysis to the $K^{(*)} {D}_1/K^{(*)} {D}_2^*$ systems, where our results suggest a series of $T_{c \bar s}$-type charmed-strange molecular tetraquark candidates. These findings provide a comprehensive picture of the molecular spectrum in the charmed-strange tetraquark sector composed of $S$-wave kaons and (anti-)charmed mesons in the $T$-doublet and can be tested in future experimental studies.
\end{abstract}
\maketitle

\section{Introduction}\label{sec1}

With the observation of a series of new hadronic states \cite{Brambilla:2019esw}, including the charmonium-like $XYZ$ states and the hidden-charm $P_c$ states, the molecular state scenario has been extensively applied to decode these novel phenomena \cite{Liu:2013waa,Hosaka:2016pey,Chen:2016qju,Richard:2016eis,Lebed:2016hpi,Liu:2019zoy,Chen:2022asf,Olsen:2017bmm,Guo:2017jvc,Meng:2022ozq,Liu:2024uxn,Wang:2025sic,Wang:2025dur,Bai:2026atm}. Among the numerous relevant studies, the heavy-flavor hadronic molecules, in particular, have attracted more attention from the community. For example, in the hidden-charm hadronic molecule interpretation, some $XYZ$ states near the thresholds of the hadron channels have been understood. Before the observation of several $P_c$ states \cite{Aaij:2019vzc}, the $P_c$-type molecular states were predicted \cite{Li:2014gra,Karliner:2015ina,Wu:2010jy,Wang:2011rga,Yang:2011wz,Wu:2012md,Chen:2015loa}, which is supported by the observed characteristic mass spectrum of the $P_c$ states \cite{Aaij:2019vzc}. In fact, the singly charmed molecular states have also been investigated \cite{Liu:2013waa,Hosaka:2016pey,Chen:2016qju,Richard:2016eis,Lebed:2016hpi,Liu:2019zoy,Chen:2022asf,Olsen:2017bmm,Guo:2017jvc,Meng:2022ozq,Liu:2024uxn,Wang:2025sic,Wang:2025dur}, especially after the discoveries of $D_{s0}(2317)$, $D_{s1}(2460)$ \cite{BaBar:2003oey,CLEO:2003ggt}, $X_0(2900)$, and $X_1(2900)$ \cite{LHCb:2020bls,LHCb:2020pxc}. Compared with studies of the hidden-charm hadronic molecules, the singly charmed molecular states have received relatively less attention, with the focus remaining on the molecular systems composed of $S$-wave kaons and $S$-wave (anti-)charmed mesons.

Based on studies of the hidden-charm molecular states in recent years, the relevant components are not limited to $S$-wave hadrons but can also include $P$-wave hadrons \cite{Liu:2013waa,Hosaka:2016pey,Chen:2016qju,Richard:2016eis,Lebed:2016hpi,Liu:2019zoy,Chen:2022asf,Olsen:2017bmm,Guo:2017jvc,Meng:2022ozq,Liu:2024uxn,Wang:2025sic,Wang:2025dur}. Considering the current status of research on the singly charmed molecular states, we note that there still exists significant potential for further investigation. Naturally, we are motivated to study the singly charmed molecular systems composed of $S$-wave kaons and $P$-wave (anti-)charmed mesons in the $T$-doublet, which is the primary focus of this work.

In this work, we first perform a dynamical study of the $T_{\bar{c}\bar{s}}$-type charmed-strange molecular tetraquark candidates formed by the $K^{(*)}$ and $\bar{D}_1/\bar{D}_2^*$ mesons\footnote{Here, $D_1$ and $D_2^*$ refer specifically to $D_1(2420)$ and $D_2^*(2460)$ \cite{ParticleDataGroup:2024cfk}.}, which is due to these configurations are unambiguously exotic due to their flavor content $\bar{c}\bar{s} q q$. We adopt the one-boson-exchange (OBE) model to derive the effective potentials of the $K^{(*)}\bar D_1/K^{(*)}\bar D_2^*$ systems, analogous to the treatment of the nuclear forces \cite{Chen:2016qju}. To ensure comprehensive and systematic results, both the $S$-$D$ wave mixing effects and the coupled-channel effects are incorporated in the realistic calculations. Using the obtained OBE effective potentials, we solve the coupled-channel Schr$\ddot{\rm o}$dinger equation to search for the loosely bound state solutions.  Additionally, we explore the $T_{{c}\bar{s}}$-type charmed-strange molecular tetraquark candidates composed of $K^{(*)}$ and $D_1/D_2^*$ mesons. The interactions in the $K^{(*)}\bar{D}_1/K^{(*)}\bar{D}_2^*$ and $K^{(*)}{D}_1/K^{(*)}{D}_2^*$ systems are connected via the $G$-parity rule \cite{Klempt:2002ap}. This investigation provides essential insights for future experimental searches for the charmed-strange molecular tetraquark candidates composed of $S$-wave kaons and $P$-wave (anti-)charmed mesons in the $T$-doublet.

This paper is organized as follows. In Sec. \ref{sec2}, we derive the effective interactions between the $K^{(*)}$ and $\bar{D}_1/\bar{D}_2^*$ mesons using the OBE model and investigate the $T_{\bar c \bar s}$-type charmed-strange molecular tetraquark candidates inspired by the $K^{(*)}\bar D_1/K^{(*)}\bar D_2^*$ interactions. In Sec. \ref{sec3}, we extend the analysis to the $T_{c \bar s}$-type charmed-strange molecular tetraquark candidates composed of $K^{(*)}$ and $D_1/D_2^*$. A summary and outlook are given in Sec. \ref{sec4}.

\section{$T_{\bar c \bar s}$-type molecular systems}\label{sec2}

In this section, we first derive the effective interactions between the $K^{(*)}$ and $\bar{D}_1/\bar{D}_2^*$ mesons using the OBE model, which has been widely employed to investigate the interactions between the hadrons \cite{Chen:2016qju}. Within this framework, the effective interactions are mediated by the exchange of light pseudoscalar, scalar, and vector mesons, such as $\pi$, $\eta$, $\sigma$, $\rho$, $\omega$, and so on, depending on the quantum numbers and the quark contents of the constituent hadrons. This framework provides an effective approach for investigating the hadronic interactions and represents a natural extension of the traditional meson-exchange picture of the nuclear force. And then, we systematically assess the possibility of the loosely bound states in the $K^{(*)}\bar{D}_1$ and $K^{(*)}\bar{D}_2^*$ systems by solving the coupled-channel Schr$\ddot{\rm o}$dinger equation with the $S$-$D$ wave mixing and coupled-channel effects included. The obtained results provide key insights into the mass spectra of the $T_{\bar c \bar s}$-type charmed-strange molecular tetraquark candidates formed by the $K^{(*)}$ and $\bar{D}_1/\bar{D}_2^*$ mesons.

We begin by classifying the charmed-strange tetraquark systems composed of the $K^{(*)}$ and $\bar{D}_1/\bar{D}_2^*$ hadrons, constructing their flavor and spin-orbital wave functions, which are essential for the subsequent calculation of the isospin factors and the operator matrix elements of the effective interactions. For the $K^{(*)}\bar{T}$ systems with $\bar T = (\bar D_1,\,\bar D_2^*)$, the flavor wave functions are constructed from the direct product of the flavor wave functions of two constituent hadrons. Since both the $K^{(*)}$ and $\bar{D}_1/\bar{D}_2^*$ mesons carry isospin with $I=1/2$, their direct product can form the charmed-strange tetraquark systems with total isospin $I=1$ or $0$. Accordingly, the flavor wave functions $|I, I_3\rangle$ for the $K^{(*)}\bar{T}$ systems are constructed as follows:
\begin{center}
\renewcommand{\arraystretch}{1.50}
\begin{tabular*}{86mm}{@{\extracolsep{\fill}}lc}
\toprule[1.00pt]
\toprule[1.00pt]
$|I, I_3\rangle$&Flavor wave functions\\
\midrule[0.75pt]
$|1,1\rangle$&$\left|K^{(*)+}{\bar T}^{0}\right\rangle$\\
$|1,0\rangle$&$\sqrt{\dfrac{1}{2}}\left|K^{(*)+}{T}^{-}\right\rangle+\sqrt{\dfrac{1}{2}}\left|K^{(*)0}{\bar T}^{0}\right\rangle$\\
$|1,-1\rangle$&$\left|K^{(*)0}{T}^{-}\right\rangle$\\
$|0,0\rangle$ &$\sqrt{\dfrac{1}{2}}\left|K^{(*)+}{T}^{-}\right\rangle-\sqrt{\dfrac{1}{2}}\left|K^{(*)0}{\bar T}^{0}\right\rangle$\\
\bottomrule[1.00pt]
\bottomrule[1.00pt]
\end{tabular*}.
\end{center}
Similar to the construction of the flavor wave functions, the spin-orbital wave functions for the $K^{(*)}\bar D_1/K^{(*)}\bar D_2^*$ systems are constructed as follows:
\begin{center}
\renewcommand{\arraystretch}{1.50}
\begin{tabular*}{86mm}{@{\extracolsep{\fill}}cc}
\toprule[1.00pt]
\toprule[1.00pt]
Systems&Spin-orbital wave functions\\
\midrule[0.75pt]
$K \bar D_1$&$\sum_{m,\,m_L}\mathbb{C}^{J,\,m_J}_{1,m;\,L,m_L}\epsilon_{m}^\mu|Y_{L,\,m_L}\rangle$\\
$K \bar D_2^*$&$\sum_{m,\,m_L}\mathbb{C}^{J,\,m_J}_{2,m;\,L,m_L}\zeta_{m}^{\mu\nu}|Y_{L,\,m_L}\rangle$\\
$K^* \bar D_1$&$\sum_{m,\,m^{\prime},\,m_S,\,m_L}\mathbb{C}^{S,\,m_S}_{1,m;\,1,m^{\prime}}\mathbb{C}^{J,\,m_J}_{S,m_S;\,L,m_L}\epsilon_{m}^\mu\epsilon_{m^{\prime}}^\nu|Y_{L,\,m_L}\rangle$\\
$K^* \bar D_2^*$&$\sum_{m,\,m^{\prime},\,m_S,\,m_L}\mathbb{C}^{S,\,m_S}_{1,m;\,2,m^{\prime}}\mathbb{C}^{J,\,m_J}_{S,m_S;\,L,m_L}\epsilon_{m}^\lambda\zeta_{m^{\prime}}^{\mu\nu}|Y_{L,\,m_L}\rangle$\\
\bottomrule[1.00pt]
\bottomrule[1.00pt]
\end{tabular*}.
\end{center}
Here, $\mathbb{C}^{e,\,f}_{a,b;\,c,d}$ is the Clebsch-Gordan coefficient, and $|Y_{L,\,m_L}\rangle$ is the spherical harmonic function. The physical quantities $\epsilon^{\mu}_m$ and $\zeta^{\mu\nu}_{m^{\prime}}$ represent the polarization vector and the polarization tensor, used to describe the polarization of the vector hadron and the tensor hadron, respectively. In the static limit, $\epsilon^{\mu}_m$ with $m=(0,\,\pm 1)$ are explicitly given by:
\begin{eqnarray}
\epsilon_{0}^{\mu}&=&\left(0,0,0,-1\right),\\
\epsilon_{\pm1}^{\mu}&=&\frac{1}{\sqrt{2}}\left(0,\,\pm 1,\,i,\,0\right).
\end{eqnarray}
$\zeta^{\mu\nu}_{m^{\prime}}$ with $m^{\prime}=(0,\,\pm 1,\,\pm 2)$ is constructed from the direct product of two polarization vectors $\epsilon^{\mu}_{m_1}$ and $\epsilon^{\nu}_{m_2}$, coupled to total angular momentum 2:
\begin{eqnarray}
\zeta^{\mu\nu}_{m^{\prime}}=\sum_{m_1,\,m_2}\mathbb{C}^{2,\,m^{\prime}}_{1,m_1;\,1,m_2}\epsilon^{\mu}_{m_1}\epsilon^{\nu}_{m_2},
\end{eqnarray}
as detailed in Ref. \cite{Cheng:2010yd}. To facilitate the following discussion, we list various components $|^{2S+1}L_J\rangle$ for the $K^{(*)}\bar D_1/K^{(*)}\bar D_2^*$ systems including the $S$-$D$ wave mixing effects:
\begin{center}
\renewcommand{\arraystretch}{1.50}
\begin{tabular*}{86mm}{@{\extracolsep{\fill}}ccccc}
\toprule[1.00pt]
\toprule[1.00pt]
$J^{P}$&$K \bar D_{1}$&$K \bar D_{2}^{*}$&$K^{*} \bar D_{1}$&$K^{*} \bar D_{2}^{*}$\\
 \midrule[0.75pt]
$0^{-}$&$\times$&$\times$&$|{}^1\mathbb{S}_{0}\rangle/|{}^5\mathbb{D}_{0}\rangle$&$\times$\\
$1^{-}$&$|{}^3\mathbb{S}_{1}\rangle/|{}^3\mathbb{D}_{1}\rangle$&$\times$&$|{}^3\mathbb{S}_{1}\rangle/|{}^{3,5}\mathbb{D}_{1}\rangle$&$|{}^3\mathbb{S}_{1}\rangle/|{}^{3,5,7}\mathbb{D}_{1}\rangle$\\
$2^{-}$&$\times$&$|{}^5\mathbb{S}_{2}\rangle/|{}^5\mathbb{D}_{2}\rangle$&$|{}^5\mathbb{S}_{2}\rangle/|{}^{1,3,5}\mathbb{D}_{2}\rangle$&$|{}^5\mathbb{S}_{2}\rangle/|{}^{3,5,7}\mathbb{D}_{2}\rangle$\\
$3^{-}$&$\times$&$\times$&$\times$&$|{}^7\mathbb{S}_{3}\rangle/|{}^{3,5,7}\mathbb{D}_{3}\rangle$\\
\bottomrule[1.00pt]
\bottomrule[1.00pt]
\end{tabular*},
\end{center}
where the symbol ``$\times$"  denotes that a channel lacks an $S$-wave component.

The construction of the OBE model begins with defining the effective Lagrangians that govern the coupling between the constituent hadrons $K^{(*)}/\bar{D}_1/\bar{D}_2^*$ and the exchanged light mesons $\mathcal{E}$ $(\mathcal{E}=\sigma/\mathbb{P}/\mathbb{V})$. In the flavor SU(3) space, the standard forms of the pseudoscalar meson matrix $\mathbb{P}$ and the vector meson matrix $\mathbb{V}_\mu$ are given as \cite{Casalbuoni:1992gi,Casalbuoni:1996pg,Yan:1992gz,Harada:2003jx,Bando:1987br}
\begin{eqnarray}
{\mathbb{P}} &=& {\left(\begin{array}{ccc}
       \frac{\pi^0}{\sqrt{2}}+\frac{\eta}{\sqrt{6}} &\pi^+ &K^+\\
       \pi^-       &-\frac{\pi^0}{\sqrt{2}}+\frac{\eta}{\sqrt{6}} &K^0\\
       K^-         &\bar K^0   &-\sqrt{\frac{2}{3}} \eta     \end{array}\right)},\\
{\mathbb{V}}_{\mu} &=& {\left(\begin{array}{ccc}
       \frac{\rho^0}{\sqrt{2}}+\frac{\omega}{\sqrt{2}} &\rho^+ &K^{*+}\\
       \rho^-       &-\frac{\rho^0}{\sqrt{2}}+\frac{\omega}{\sqrt{2}} &K^{*0}\\
       K^{*-}         &\bar K^{*0}   & \phi     \end{array}\right)}_{\mu},
\end{eqnarray}
respectively.

The effective Lagrangians describing the anticharmed mesons in $T$-doublet $\bar{D}_1/\bar{D}_2^*$ coupling with the exchanged light mesons $\mathcal{E}$ are built using the heavy quark symmetry, the chiral symmetry, and the hidden local symmetry \cite{Casalbuoni:1992gi,Casalbuoni:1996pg,Yan:1992gz,Harada:2003jx,Bando:1987br}. They are given as follows \cite{Ding:2008gr}:
\begin{eqnarray}\label{effectiveLagrangians1}
\mathcal{L}_{\bar{T}\bar{T}\mathcal{E}}&=&-2g_\sigma^{\prime\prime}\bar{D}_{1a\mu}\bar{D}^{\mu\dagger}_{1a}\sigma
+2g_\sigma^{\prime\prime}\bar{D}^{*\dagger}_{2a\mu\nu}\bar{D}^{*\mu\nu}_{2a}\sigma\nonumber\\
    &&-\frac{5ik}{3f_\pi}\varepsilon^{\mu\nu\rho\tau}v_\nu\bar{D}^{\dag}_{1a\rho}
\bar{D}_{1b\tau}\partial_\mu\mathbb{P}_{ba}\nonumber\\
    &&+\frac{2ik}{f_\pi}\varepsilon^{\mu\nu\rho\tau}v_\nu\bar{D}^{*\alpha\dag}_{2a\rho}
    \bar{D}^{*}_{2b\alpha\tau}\partial_\mu\mathbb{P}_{ba}\nonumber\\
    &&+\sqrt{\frac{2}{3}}\frac{k}{f_\pi}\left(\bar{D}^{\dagger}_{1a\mu}
    \bar{D}^{*\mu\lambda}_{2b}+\bar{D}_{1b\mu}
    \bar{D}^{*\mu\lambda\dagger}_{2a}\right)\partial_\lambda\mathbb{P}_{ba}\nonumber\\
    &&+\sqrt{2}\beta^{\prime\prime}g_{V}\left(v\cdot\mathbb{V}_{ba}\right)\bar{D}_{1b\mu}
\bar{D}^{\mu\dagger}_{1a}\nonumber\\
    &&+\frac{5\sqrt{2}i\lambda^{\prime\prime}g_{V}}{3}\left(\bar{D}^{\nu}_{1b}
    \bar{D}^{\mu\dagger}_{1a}
    -\bar{D}^{\nu\dagger}_{1a}\bar{D}^{\mu}_{1b}\right)\partial_\mu\mathbb{V}_{ba\nu}\nonumber\\
    &&-\sqrt{2}\beta^{\prime \prime}g_{V}\left(v\cdot\mathbb{V}_{ba}\right) \bar{D}_{2b}^{*\lambda\nu} \bar{D}^{*\dagger}_{2a{\lambda\nu}}\nonumber\\
    &&+2\sqrt{2}i\lambda^{\prime\prime} g_{V}\left(\bar{D}^{*\lambda\nu\dagger}_{2a}
    \bar{D}^{*\mu}_{2b\lambda}-\bar{D}^{*\lambda\nu}_{2b} \bar{D}^{*\mu\dagger}_{2a\lambda}\right)\partial_\mu \mathbb{V}_{ba\nu}\nonumber\\
    &&+\frac{i\beta^{\prime\prime}g_{V}}{\sqrt{3}}\varepsilon^{\lambda\alpha\rho\tau}v_{\rho}
    \left(v\cdot\mathbb{V}_{ba}\right)\left(\bar{D}^{\dagger}_{1a\alpha} \bar{D}^{*}_{2b\lambda\tau}-\bar{D}_{1b\alpha}
    \bar{D}^{\dagger*}_{2a\lambda\tau}\right)\nonumber\\
    &&+\frac{2\lambda^{\prime\prime}g_{V}}{\sqrt{3}}
    \left[3\varepsilon^{\mu\lambda\nu\tau}v_\lambda\left(\bar{D}^{\alpha\dagger}_{1a}
    \bar{D}^{*}_{2b\alpha\tau}+\bar{D}^{\alpha}_{1b}
    \bar{D}^{*\dagger}_{2a\alpha\tau}\right)\partial_\mu\mathbb{V}_{ba\nu}\right.\nonumber\\
    &&\left.+2\varepsilon^{\lambda\alpha\rho\nu}v_\rho
    \left(\bar{D}^{\dagger}_{1a\alpha}\bar{D}^{*\mu}_{2b\lambda}
    +\bar{D}_{1b\alpha}\bar{D}^{\dagger\mu*}_{2a\lambda}\right)\right.\nonumber\\
    &&\left.\times\left(\partial_\mu \mathbb{V}_{ba\nu}-\partial_\nu \mathbb{V}_{ba\mu}\right)\right].
\end{eqnarray}
Here, $\bar D_1=(\bar D_1^{0},\,D_1^{-})$, $\bar D_2^*=(\bar D_2^{*0},\,D_2^{*-})$, and the normalization relations for the anticharmed mesons in $T$-doublet are \cite{Ding:2008gr}
\begin{eqnarray}
\langle 0|\bar D_{1}^{\mu}|\bar {c} q(1^+)\rangle&=&\sqrt{m_{D_{1}}}\epsilon^\mu,\\
\langle 0|\bar D_{2}^{*\mu\nu}|\bar {c} q(2^+)\rangle&=&\sqrt{m_{D_{2}^*}}\zeta^{\mu\nu}.
\end{eqnarray}

Following Refs. \cite{Liu:2020nil,Wang:2024ukc,Wang:2024kke}, the effective Lagrangians for the couplings of the $K^{(*)}$ mesons to the exchanged light mesons $\mathcal{E}$ are:
\begin{eqnarray}\label{effectiveLagrangians2}
\mathcal{L}_{K^{(*)}K^{(*)}\mathcal{E}} &=&-2g_{\sigma}{K}_a {K}_a^{\dag} \sigma+ 2g_{\sigma} {K}_{a\mu}^* {K}_a^{*\mu\dag} \sigma\nonumber\\
    &&+\frac{2ig}{f_{\pi}}v^{\alpha}\varepsilon_{\alpha\mu\nu\lambda}{K}_a^{*\mu\dag}{K}_b^{*\lambda}\partial^{\nu}{\mathbb{P}}_{ab}\nonumber\\
    &&+\frac{2g}{f_{\pi}}\left({K}_a^{*\mu\dag}{K}_b+{K}_a^{\dag}{K}_b^{*\mu}\right)\partial_{\mu}{\mathbb{P}}_{ab}\nonumber\\
    &&+\sqrt{2}\beta g_V {K}_a {K}_b^{\dag} v\cdot\mathbb{V}_{ab}-\sqrt{2}\beta g_V {K}_{a\mu}^* {K}_b^{*\mu\dag}v\cdot\mathbb{V}_{ab}\nonumber\\
    &&-2\sqrt{2}i\lambda g_V {K}_a^{*\mu\dag}{K}_b^{*\nu}\left(\partial_{\mu}\mathbb{V}_{\nu}-\partial_{\nu}\mathbb{V}_{\mu}\right)_{ab}\nonumber\\
    &&-2\sqrt{2}\lambda g_V v^{\lambda}\varepsilon_{\lambda\mu\alpha\beta}\left({K}_a^{*\mu\dag}{K}_b+{K}_a^{\dag}{K}_b^{*\mu}\right)\partial^{\alpha}\mathbb{V}^{\beta}_{ab},\nonumber\\
\end{eqnarray}
where $K^{(*)}=(K^{(*)+},\,K^{(*)0})$, and the normalization relations for the $K^{(*)}$ mesons are
\begin{eqnarray}
\langle 0|K|q\bar{s}(0^-)\rangle&=&\sqrt{m_{K}},\\
\langle 0|K^{*\mu}|q\bar{s}(1^-)\rangle&=&\sqrt{m_{K^*}}\epsilon^\mu.
\end{eqnarray}

The coupling constants of the effective Lagrangians in Eqs.~(\ref{effectiveLagrangians1}) and (\ref{effectiveLagrangians2}) parameterize the effective interaction strengths of the $\bar T \bar T \mathcal{E}$ and $K^{(*)} K^{(*)} \mathcal{E}$ vertices, which are constrained by the experimental data or the theoretical models. Specifically, the coupling constants governing the interactions between the anticharmed mesons in $T$-doublet and the exchanged light mesons can be derived from the quark model \cite{Riska:2000gd}, and we use $g_{\sigma}^{\prime\prime}=0.76$, $k=0.59$, $\beta^{\prime\prime}=0.90$, and $\lambda^{\prime\prime}=0.56 ~\rm {GeV^{-1}}$ in this work \cite{Wang:2019aoc,Wang:2020dya,Wang:2019nwt,Wang:2021yld,Wang:2021aql,Yang:2021sue,Wang:2021ajy,Wang:2023ftp,Wang:2023ivd}. The value $g=-1.12$ \cite{Wang:2024ukc,Wang:2024kke} is extracted from the measured width of the $K^* \to K \pi$ process \cite{ParticleDataGroup:2024cfk}.  The parameters $g_{\sigma}=-2.82$ \cite{Sheng:2024hkf,Chen:2022svh} and $\lambda=-0.56$ GeV$^{-1}$ \cite{Liu:2020nil,Wang:2024ukc,Wang:2024kke} are also obtained from the quark model \cite{Riska:2000gd}, while  $\beta=-0.835$ follows from the hidden-gauge symmetry of the vector meson arguments \cite{Molina:2010tx}. Additionally, we take $f_{\pi}=132$ MeV and $g_V=5.83$ \cite{Bando:1987br,Isola:2003fh}. Here, we need to mention that the relative phase factors of the related coupling constants are fixed consistently using the quark model \cite{Riska:2000gd}.

After establishing the effective Lagrangians and corresponding coupling constants describing the $\bar T \bar T \mathcal{E}$ and $K^{(*)} K^{(*)}{\mathcal{E}}$ interaction vertices, we now derive the effective interactions in the coordinate space between the $K^{(*)}$ and $\bar{D}_1/\bar{D}_2^*$ hadrons within the OBE model. The effective interactions for the $K^{(*)}\bar D_1/K^{(*)}\bar D_2^*$ systems are derived by calculating the corresponding scattering amplitudes for the processes $K^{(*)}\bar T \to K^{(*)}\bar T$ via the exchange of a serious of light mesons ${\mathcal{E}}$ (${\mathcal{E}}=\sigma,\,\pi,\,\eta,\,\rho,\,\omega$) \cite{Chen:2016qju}. Applying the Feynman rules to the interaction vertices defined by the effective Lagrangians, the scattering amplitude for the process $K^{(*)}\bar T \to K^{(*)}\bar T$ via the exchange of a light meson ${\mathcal{E}}$ $\mathcal{M}_{{\mathcal{E}}}^{K^{(*)}\bar T \to K^{(*)}\bar T}(\bm{q})$ can be obtained. And then, the effective interaction in the momentum space for the $K^{(*)}\bar T \to K^{(*)}\bar T$ process ${V}_{{\mathcal{E}}}^{K^{(*)}\bar T \to K^{(*)}\bar T}(\bm{q})$ is related to the corresponding scattering amplitude:
\begin{eqnarray}
{V}_{{\mathcal{E}}}^{K^{(*)}\bar T \to K^{(*)}\bar T}(\bm{q})=-\frac{\mathcal{M}_{{\mathcal{E}}}^{K^{(*)}\bar T \to K^{(*)}\bar T}(\bm{q})} {4\sqrt{m_{K^{(*)}}m_{\bar T}m_{K^{(*)}}m_{\bar T}}},
\end{eqnarray}
which follows from the Breit approximation \cite{Berestetskii:1982qgu}. Finally, the effective interaction in the coordinate space for the $K^{(*)}\bar T \to K^{(*)}\bar T$ process ${V}_{{\mathcal{E}}}^{K^{(*)}\bar T \to K^{(*)}\bar T}(\bm{r})$ is obtained by performing the Fourier transformation:
\begin{eqnarray}\label{FT}
{V}_{{\mathcal{E}}}^{K^{(*)}\bar T \to K^{(*)}\bar T}(\bm{r})=\int\frac{d^3\bm{q}}{(2\pi)^3}e^{i\bm{q}\cdot\bm{r}}{V}_{{\mathcal{E}}}^{K^{(*)}\bar T \to K^{(*)}\bar T}(\bm{q}){F}_{\rm M}^2(q,\,m_{{\mathcal{E}}}).\nonumber\\
\end{eqnarray}
To account for the finite size of the hadrons and to regularize the high-energy behavior of the resulting interaction, the vertex form factor is introduced in the Eq. (\ref{FT}). A common choice is the monopole form factor ${F}_{\rm M}(q,\,m_{{\mathcal{E}}})$ when studying the bound state properties of the molecular states \cite{Tornqvist:1993ng,Tornqvist:1993vu}, i.e.,
\begin{eqnarray}
{F}_{\rm M}(q,\,m_{{\mathcal{E}}}) = \frac{\Lambda^2-m_{{\mathcal{E}}}^2}{\Lambda^2-q^2},
\end{eqnarray}
where $\Lambda$ is a cutoff parameter. Within the OBE model, the cutoff is a free parameter. In the absence of direct experimental constraints, its precise value cannot be determined unambiguously. Thus, we vary $\Lambda$ in our numerical analysis to explore the emergence of  the loosely bound states. Following the practice established in studies of the molecular candidates such as the deuteron, $P_{c}$, and $T_{cc}$, a cutoff scale of order 1 GeV that produces a loosely bound state is widely adopted as a physically reasonable input. Consequently, the loosely bound state with the cutoff value closed to 1 GeV is more likely to regarded as the possible hadronic molecular candidate \cite{Chen:2016qju}.

The total OBE effective interaction in the coordinate space for the $K^{(*)}\bar T \to K^{(*)}\bar T$ process ${V}^{K^{(*)}\bar T \to K^{(*)}\bar T}(\bm{r})$ is the sum of the contributions from all considered meson exchange interactions:
\begin{eqnarray}
{V}^{K^{(*)}\bar T \to K^{(*)}\bar T}(\bm{r})=\sum_{{\mathcal{E}}=\sigma,\,\pi,\,\eta,\,\rho,\,\omega}{V}_{{\mathcal{E}}}^{K^{(*)}\bar T \to K^{(*)}\bar T}(\bm{r}).
\end{eqnarray}
In Appendix \ref{app01}, we summarise the obtained OBE effective interactions in the coordinate space for the $K^{(*)}\bar D_1/K^{(*)}\bar D_2^*$ systems.

To determine whether the $K^{(*)}\bar D_1/K^{(*)}\bar D_2^*$ systems  can serve as the candidates for the hadronic molecular states, we solve the coupled-channel Schr$\ddot{\rm o}$dinger equation with the OBE effective interactions in the coordinate space ${V}^{K^{(*)}\bar T \to K^{(*)}\bar T}(\bm{r})$. The numerical analysis entails scanning over the cutoff parameter $\Lambda$ to search for the loosely bound state solutions for the $K^{(*)}\bar D_1/K^{(*)}\bar D_2^*$ systems, from which we extract the binding energy and the spatial wave functions of various components. The obtained results provide critical insight into the viability of the $K^{(*)}\bar D_1/K^{(*)}\bar D_2^*$ systems as the $T_{\bar c \bar s}$-type charmed-strange molecular tetraquark candidates. Here, we need to mention that the loosely bound state can be considered as a promising molecular candidate if it is weakly bound (small binding energy) and spatially extended (large RMS radius) for the cutoff parameter closed to 1.0 GeV \cite{Chen:2016qju}, consistent with established interpretations of the hadronic molecules for many observed new hadrons \cite{Liu:2013waa,Hosaka:2016pey,Chen:2016qju,Richard:2016eis,Lebed:2016hpi,Liu:2019zoy,Chen:2022asf,Olsen:2017bmm,Guo:2017jvc,Meng:2022ozq,Liu:2024uxn,Wang:2025sic,Wang:2025dur}.

\begin{figure}[htbp]
  \centering
  \includegraphics[width=8.6cm]{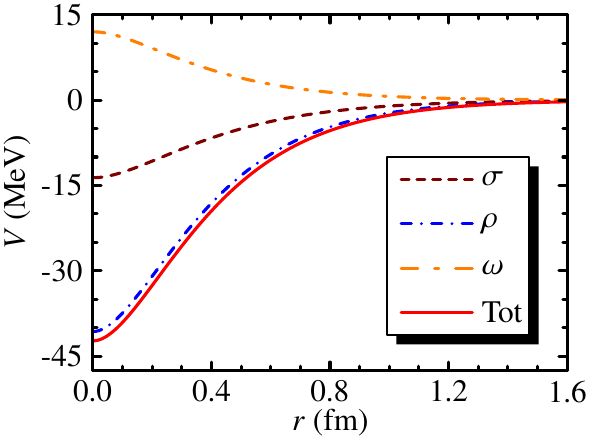}
  \caption{Effective interaction for the $K\bar{D}_1$ system with $I(J^P)=0(1^-)$ with $\Lambda=1.0~{\rm GeV}$.}\label{KbarD1potential}
\end{figure}

First, we perform a single channel analysis. In the single-channel analysis, we observe that  the $K^* \bar D_1$ states with $I(J^P)=0(0^-,\,1^-)$ and the $K^* \bar D_2^*$ states with $I(J^P)=0(1^-,\,2^-)$ can form the loosely bound states  when the cutoff parameters fall within a reasonable range, whereas the other systems under consideration do not yield the loosely bound state solutions. We then proceed to present a detailed discussion. For the $K \bar D_1$ system with $I(J^P)=0(1^-)$, we first analyze its effective interaction. In Fig.~\ref{KbarD1potential}, we present the effective interaction for the $K\bar{D}_1$ system with $I(J^P)=0(1^-)$ with $\Lambda=1.0\,{\rm GeV}$. For a given hadron-hadron system, the allowed exchanged mesons should satisfy the spin-parity conservation. Since the $KK\pi$ vertex is strictly forbidden by the spin-parity conservation, there only exists the $\sigma$, $\rho$, and $\omega$ exchange contributions to the $K\bar{D}_1$ system. Our analysis indicates that the $\sigma$ exchange yields an attractive interaction, while the $\rho$ exchange also contributes attractively. In contrast, the $\omega$ exchange gives rise to a repulsive interaction. It is noteworthy that the attractive interaction from the $\sigma$ exchange and the repulsive interaction from the $\omega$ exchange largely cancel each other. Consequently, the total interaction is dominated by the attractive contribution from the $\rho$ exchange interaction. Since the quark content of the $K \bar D_1$ system is $q\bar s$-$\bar c q$, previous theoretical work has established that the $\sigma$ exchange generally provides attraction and the $\omega$ exchange induces repulsion in such system \cite{Chen:2017vai}, a conclusion is consistent with our results obtained within the OBE model. We next perform a numerical calculation to search for the loosely bound state solutions for the $K \bar D_1$ system with $I(J^P)=0(1^-)$. In both the single-channel analysis and the $S$-$D$ wave mixing analysis, we find that no loosely bound state emerges in the $K \bar D_1$ system with $I(J^P)=0(1^-)$ for the cutoff parameters ranging from 0.8 to 2.0 GeV. Moreover, the $D$-wave channel contribution is negligible, owing to the absence of the tensor-force interaction in the $K \bar D_1$ system. Therefore, based on both the single-channel analysis and the $S$-$D$ wave mixing analysis, the $K \bar D_1$ system with $I(J^P)=0(1^-)$ does not form a loosely bound state. This conclusion aligns with the majority of existing theoretical studies \cite{Dong:2020rgs,He:2020btl}.

Analogous to the OBE effective interaction of the $K \bar D_1$ system with $I(J^P)=0(1^-)$, the OBE effective interaction of the $K \bar D_2^*$ system with $I(J^P)=0(2^-)$ is mediated only by the $\sigma$, $\rho$, and $\omega$ exchanges. Among these, the $\sigma$ and $\rho$ exchange interactions provide attraction, and the $\omega$ exchange interaction provides repulsion. Similar to the binding behavior of the $K \bar D_1$ system with $I(J^P)=0(1^-)$, the single-channel analysis and the $S$-$D$ wave mixed analysis, with a cutoff parameter $\Lambda$ varying between 0.8 and 2.0 GeV, show that the $K \bar D_2^*$ system with $I(J^P)=0(2^-)$ does not support the formation of a loosely bound state. Thus, the $K \bar D_2^*$ system with $I(J^P)=0(2^-)$ does not form a loosely bound state in the single-channel analysis and the $S$-$D$ wave mixed analysis.

\renewcommand\tabcolsep{0.27cm}
\renewcommand{\arraystretch}{1.50}
\begin{table}[!htbp]
\caption{The obtained loosely bound state solutions for the $K^{*} \bar D_1$ system with a cutoff parameter $\Lambda$ varying between 0.8 and 2.0 GeV. The $P$ are probabilities of various components (in units of \%).}\label{KstbarD1}
\begin{tabular*}{86mm}{@{\extracolsep{\fill}}ccccc}
\toprule[1pt]
\toprule[1pt]
\multirow{10}{*}{$0(0^{-})$}&\multicolumn{4}{c}{Single channel case}\\
\Xcline{2-5}{0.75pt}
&$\Lambda~(\rm{GeV})$ &$E~(\rm {MeV})$ &$r_{\rm RMS}~(\rm {fm})$\\
&$0.97$&$-0.41$&$4.90$\\
&$1.04$&$-6.51$&$1.67$\\
&$1.11$&$-20.35$&$1.03$\\
\Xcline{2-5}{0.75pt}
&\multicolumn{4}{c}{$S$-$D$ wave mixing case}\\
\Xcline{2-5}{0.75pt}
&$\Lambda~(\rm{GeV})$ &$E~(\rm {MeV})$ &$r_{\rm RMS}~(\rm {fm})$& $P$(${}^1\mathbb{S}_{0}/{}^5\mathbb{D}_{0}$)\\
&$0.95$&$-0.51$&$4.69$&\textbf{99.59}/0.41\\
&$1.03$&$-7.63$&$1.59$&\textbf{99.29}/0.71\\
&$1.10$&$-21.62$&$1.03$&\textbf{99.28}/0.72\\\midrule[1.0pt]
\multirow{15}{*}{$0(1^{-})$}&\multicolumn{4}{c}{Single channel case}\\
\Xcline{2-5}{0.75pt}
&$\Lambda~(\rm{GeV})$ &$E~(\rm {MeV})$ &$r_{\rm RMS}~(\rm {fm})$\\
&$1.29$&$-0.40$&$4.95$\\
&$1.39$&$-6.88$&$1.63$\\
&$1.48$&$-19.63$&$1.03$\\
\Xcline{2-5}{0.75pt}
&\multicolumn{4}{c}{$S$-$D$ wave mixing case}\\
\Xcline{2-5}{0.75pt}
&$\Lambda~(\rm{GeV})$ &$E~(\rm {MeV})$ &$r_{\rm RMS}~(\rm {fm})$& $P$(${}^3\mathbb{S}_{1}/{}^3\mathbb{D}_{1}/{}^5\mathbb{D}_{1}$)\\
&$1.22$&$-0.33$&$5.21$&\textbf{99.41}/0.59/$\mathcal{O}(0)$\\
&$1.33$&$-6.81$&$1.68$&\textbf{98.70}/1.30/$\mathcal{O}(0)$\\
&$1.43$&$-20.39$&$1.05$&\textbf{98.52}/1.48/$\mathcal{O}(0)$\\
\bottomrule[1pt]\bottomrule[1pt]
\end{tabular*}
\end{table}

Table \ref{KstbarD1} summarizes the loosely bound state solutions obtained for the $K^{*} \bar D_1$ system with a cutoff parameter $\Lambda$ varying between 0.8 and 2.0 GeV. In the case of the $K^* \bar D_1$ system with $I(J^P)=0(0^-)$, the single-channel analysis and the $S$-$D$ wave mixing analysis yield nearly identical results, indicating that the $S$-$D$ wave mixing effects have a negligible impact on the formation of this bound state. In the single-channel analysis, a loosely bound state emerges at a cutoff parameter of 0.97 GeV. In the $S$-$D$ wave mixing analysis, a loosely bound state is formed at a cutoff of 0.95 GeV, with the $S$-wave component accounting for more than 99\% of the spatial wave function. In the case of the $K^* \bar D_1$ system with $I(J^P)=0(1^-)$, a loosely bound state is found in the single-channel analysis at a cutoff parameter of 1.29 GeV. When the $D$-wave contribution is included, the $K^* \bar D_1$ system with $I(J^P)=0(1^-)$ still forms a loosely bound state at a slightly lower cutoff of 1.22 GeV, with the $S$-wave component remaining dominant at over 98\%. 

\renewcommand\tabcolsep{0.20cm}
\renewcommand{\arraystretch}{1.50}
\begin{table}[htbp]
\caption{The obtained loosely bound state solutions for the $K^{*} \bar D_2^{*}$ system with a cutoff parameter $\Lambda$ varying between 0.8 and 2.0 GeV. The $P$ are probabilities of various components (in units of \%).}\label{KstbarD2st}
\begin{tabular*}{88mm}{@{\extracolsep{\fill}}ccccc}\toprule[1pt]\toprule[1pt]
\multirow{10}{*}{$0(1^{-})$}&\multicolumn{4}{c}{Single channel case}\\
\Xcline{2-5}{0.75pt}
&$\Lambda~(\rm{GeV})$ &$E~(\rm {MeV})$ &$r_{\rm RMS}~(\rm {fm})$\\
&$1.02$&$-0.66$&$4.34$\\
&$1.09$&$-7.08$&$1.61$\\
&$1.16$&$-20.74$&$1.02$\\
\Xcline{2-5}{0.75pt}
&\multicolumn{4}{c}{$S$-$D$ wave mixing case}\\
\Xcline{2-5}{0.75pt}
&$\Lambda~(\rm{GeV})$ &$E~(\rm {MeV})$ &$r_{\rm RMS}~(\rm {fm})$& $P$(${}^3\mathbb{S}_{1}/{}^3\mathbb{D}_{1}/{}^5\mathbb{D}_{1}/{}^7\mathbb{D}_{1}$)\\
&$0.98$&$-0.47$&$4.79$&\textbf{99.41}/0.06/0.41/0.12\\
&$1.06$&$-6.93$&$1.66$&\textbf{98.92}/0.11/0.76/0.22\\
&$1.14$&$-21.87$&$1.02$&\textbf{98.87}/0.12/0.80/0.22\\\midrule[1.0pt]
\multirow{10}{*}{$0(2^{-})$}&\multicolumn{4}{c}{Single channel case}\\
\Xcline{2-5}{0.75pt}
&$\Lambda~(\rm{GeV})$ &$E~(\rm {MeV})$ &$r_{\rm RMS}~(\rm {fm})$\\
&$1.56$&$-0.40$&$4.98$\\
&$1.70$&$-6.65$&$1.65$\\
&$1.84$&$-20.77$&$1.00$\\
\Xcline{2-5}{0.75pt}
&\multicolumn{4}{c}{$S$-$D$ wave mixing case}\\
\Xcline{2-5}{0.75pt}
&$\Lambda~(\rm{GeV})$ &$E~(\rm {MeV})$ &$r_{\rm RMS}~(\rm {fm})$& $P$(${}^5\mathbb{S}_{2}/{}^3\mathbb{D}_{2}/{}^5\mathbb{D}_{2}/{}^7\mathbb{D}_{2}$)\\
&$1.40$&$-0.50$&$4.79$&\textbf{99.03}/0.37/0.44/0.16\\
&$1.54$&$-6.62$&$1.72$&\textbf{97.92}/0.80/0.94/0.34\\
&$1.68$&$-20.29$&$1.06$&\textbf{97.44}/0.99/1.16/0.40\\
\bottomrule[1pt]\bottomrule[1pt]
\end{tabular*}
\end{table}

In Table \ref{KstbarD2st}, we present the obtained loosely bound state solutions for the $K^{*} \bar D_2^{*}$ system with a cutoff parameter $\Lambda$ varying between 0.8 and 2.0 GeV. For the $K^* \bar D_2^*$ system with $I(J^P)=0(1^-)$, the single-channel analysis yields a loosely bound state solution at $\Lambda=1.02\,{\rm GeV}$, with a binding energy of 0.66 MeV and a RMS radius of 4.34 fm. When the cutoff is increased to 1.16 GeV, the binding energy reaches 20.74 MeV, corresponding to a RMS radius of approximately 1.02 fm. We further examine the effects of the $S$-$D$ wave mixing. A loosely bound state is found at $\Lambda=0.98\,{\rm GeV}$, with the $S$-wave component dominating its composition. Specifically, for the binding energies below 20 MeV, the $S$-wave probability exceeds 98\%. For the $K^* \bar D_2^*$ system with $I(J^P)=0(2^-)$, a single-channel analysis indicates the formation of a loosely bound state at $\Lambda=1.56\,{\rm GeV}$, where the corresponding binding energy increases monotonically with the cutoff parameter. When the $D$-wave contribution is included, a loosely bound state emerges at a lower cutoff of $\Lambda=1.40\,{\rm GeV}$, while the $S$-wave remains the dominant component. 

In recent years, the coupled-channel analysis have been applied to study the hadronic molecules \cite{Chen:2016qju}. In this work, we perform a coupled-channel analysis of some focused systems.

\begin{figure}[htbp]
  \centering
  \includegraphics[width=8.6cm]{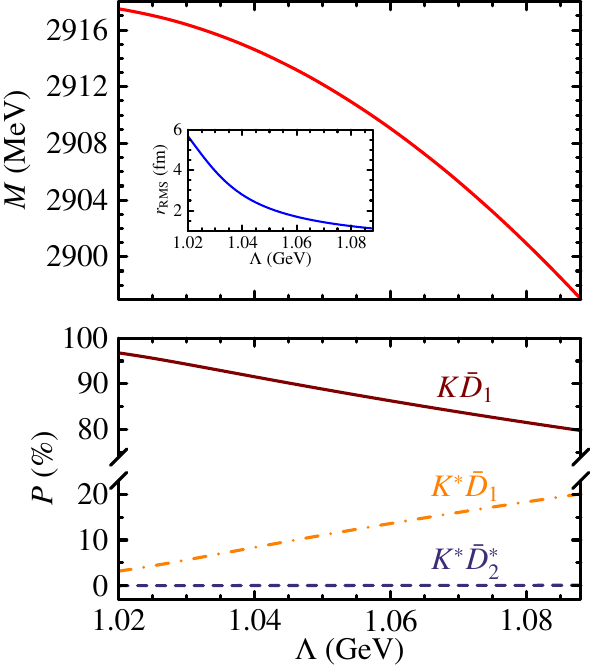}
  \caption{The bound state properties including the mass, the RMS radius, and the probabilities of various components as functions of the cutoff $\Lambda$ for the coupled $K \bar D_1 / K^* \bar D_1 / K^* \bar D_2^*$ system with $I(J^P)=0(1^-)$.}\label{KbarD1}
\end{figure}

For the coupled $K \bar D_1 / K^* \bar D_1 / K^* \bar D_2^*$ system with $I(J^P)=0(1^-)$, we can obtain the loosely bound state solutions. Fig.~\ref{KbarD1} shows the bound state properties including the mass, the RMS radius, and the probabilities of various components as functions of the cutoff $\Lambda$ for the coupled $K \bar D_1 / K^* \bar D_1 / K^* \bar D_2^*$ system with $I(J^P)=0(1^-)$. We find that when the cutoff parameter is around 1.02 GeV, the coupled $K \bar D_1 / K^* \bar D_1 / K^* \bar D_2^*$ system with $I(J^P)=0(1^-)$ can form a loosely bound state. Notably, besides the $K\bar{D}_1$ component, the $K^*\bar{D}_1$ channel also contributes significantly. In fact, the coupled $K \bar D_1 / K^* \bar D_1 / K^* \bar D_2^*$ system with $I(J^P)=0(1^-)$ may be related to the observed $X_1(2900)$. At a cutoff parameter of $\Lambda \approx 1.07\,\rm{GeV}$, the mass of the coupled $K \bar D_1 / K^* \bar D_1 / K^* \bar D_2^*$ bound state with $I(J^P)=0(1^-)$ is calculated to be 2904 MeV. The corresponding RMS radius is determined to be 1.38 fm, and an analysis of the channel components reveals the following probabilities: the $K \bar D_1 $ channel constitutes about 83.41\%, the $K^* \bar D_1 $ channel contributes 19.50\%, and the $K^* \bar D_2^* $ channel has a minimal presence of approximately 0.09\%.

\begin{figure}[htbp]
  \centering
  \includegraphics[width=8.6cm]{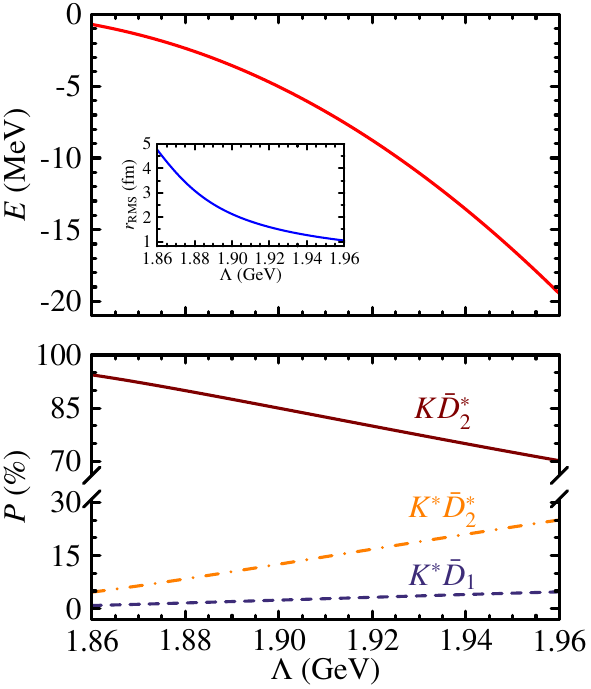}
  \caption{The binding energy, the RMS radius, and the probabilities of various components as functions of the cutoff $\Lambda$ for the coupled $K \bar D_2^* / K^* \bar D_1 / K^* \bar D_2^*$ system with $I(J^P)=0(2^-)$.}\label{KbarD2}
\end{figure}

For the coupled $K \bar D_2^* / K^* \bar D_1 / K^* \bar D_2^*$ system with $I(J^P)=0(2^-)$, we find that the loosely bound state solutions can be obtained. In Fig.~\ref{KbarD2}, we present the resulting bound state properties including the binding energy, the RMS radius, and the probabilities of various components as functions of the cutoff $\Lambda$ for the coupled $K \bar D_2^* / K^* \bar D_1 / K^* \bar D_2^*$ system with $I(J^P)=0(2^-)$. When the cutoff parameter is fixed to be larger than $1.86\,{\rm GeV}$, a loosely bound state is predicted. Notably, besides the dominant $K \bar D_2^*$ component, the $K^* \bar D_2^*$ channel also plays an important role in this bound state.

Based on the numerical results obtained, we find that the $K\bar{D}_1$ system with $I(J^P)=0(1^-)$ and the $K \bar D_2^*$ system with $I(J^P)=0(2^-)$ do not support the loosely bound state solutions in the single-channel analysis, whereas such solutions can emerge when the coupled-channel effects are taken into account. This reveals a deviation between the numerical results and the expected physical picture, wherein channels far away from the system under consideration are generally believed to play only a minor role in refining the properties of the dominant channel\footnote{In the discussed $K \bar D_1 / K^* \bar D_1 / K^* \bar D_2^*$ coupled-channel system with $I(J^P)=0(1^-)$, the second
threshold is too high above the first threshold ($\sim 400$ MeV).}. Thus, we analyze the possible reason in the following. We check the off-diagonal potential of this coupled $K \bar D_1 / K^* \bar D_1 / K^* \bar D_2^*$ system with $I(J^P)=0(1^-)$ and find that the pion exchange occurs for the $K \bar D_1 - K^* \bar D_1$ coupling\footnote{When the strength of the $K\bar{D}_1 \to K^*\bar{D}_1$ coupling is reduced by 80\%, the bound state solution in the coupled $K\bar{D}_1/K^*\bar{D}_1/K^*\bar{D}_2^*$ system with $I(J^P)=0(1^-)$ disappears.}. In our calculation, we usually take the same cutoff value when deriving the effective potential of the coupled-channel system to avoid involving more model parameters. In fact, this treatment results in a large off-diagonal contribution for the $K \bar D_1 -K^* \bar D_1$ coupling. In our previous work \cite{Wang:2022mxy}, we encountered similar situations where the coupled-channel effects altered the conclusions obtained from the single-channel analyses. We attributed this primarily to the use of a common cutoff parameter for different channels in the coupled-channel calculations, and this issue can be resolved by assigning different cutoff parameters to different channels. However, due to the absence of experimental data for the systems under investigation in this work, we adopted the same cutoff parameter for the related channels in order to minimize the number of free parameters.

The same situation occurs for the coupled $K \bar D_2^* / K^* \bar D_1 / K^* \bar D_2^*$ system with $I(J^P)=0(2^-)$ discussed above.

Consequently, caution should be exercised when concluding that the loosely bound states exist in the coupled $K \bar D_1 / K^* \bar D_1 / K^* \bar D_2^*$ system with $I(J^P)=0(1^-)$ and the coupled $K \bar D_2^* / K^* \bar D_1 / K^* \bar D_2^*$ system with $I(J^P)=0(2^-)$.

Considering the above analysis, we suggest that the $K^* \bar D_1$ states with $I(J^P)=0(0^-,\,1^-)$ and the $K^* \bar D_2^*$ states with $I(J^P)=0(1^-,2^-)$ represent the most promising candidates of the $T_{\bar{c}\bar{s}}$-type charmed-strange molecular tetraquarks, while the coupled $K \bar D_1 / K^* \bar D_1 / K^* \bar D_2^*$ system with $I(J^P)=0(1^-)$ and the coupled $K \bar D_2^* / K^* \bar D_1 / K^* \bar D_2^*$ system with $I(J^P)=0(2^-)$ can only be regarded as possible candidates of the $T_{\bar{c}\bar{s}}$-type charmed-strange molecular tetraquarks. In view of the limited experimental resources, we suggest that priority be given in future searches to the $K^* \bar D_1$ molecular states with $I(J^P)=0(0^-,1^-)$ and the $K^* \bar D_2^*$ molecular states with $I(J^P)=0(1^-,2^-)$. 

Finally, we discuss whether the $X_1(2900)$ can be interpreted as a charmed-strange molecular state in light of our results. Based solely on the single-channel analysis in this section, our results do not support the interpretation of the $X_1(2900)$ as a $K \bar D_1$ molecule with $I(J^P)=0(1^-)$\footnote{In Refs.~\cite{Qi:2021iyv}, the $X_1(2900)$, reported by LHCb in the $B^+\to D^+D^-K^+$ process~\cite{LHCb:2020bls,LHCb:2020pxc}, was suggested as a charmed-strange molecular state.}. Since there exists uncertainty in the coupled-channel analysis of the $K\bar{D}_1/K^*\bar{D}_1/K^*\bar{D}_2^*$ system with $I(J^P)=0(1^-)$, caution should be exercised when concluding that the $X_1(2900)$ is a charmed-strange hadronic molecular state.

\section{$T_{c \bar s}$-type molecular systems}\label{sec3}

In the previous section,  we presented predictions for the $T_{\bar c \bar s}$-type charmed-strange molecular tetraquark candidates formed by the $K^{(*)}$ and $\bar{D}_1/\bar{D}_2^*$ mesons. In the present section, we extend our investigation to the $T_{c \bar s}$-type charmed-strange molecular tetraquark candidates composed of the $K^{(*)}$ and $D_1/D_2^*$ mesons. It is also noteworthy that in Ref. \cite{Guo:2011dd}, the $K {D}_1$ and $K {D}_2^*$ systems were investigated as the hadronic molecules using the chiral and heavy quark symmetries.

A key element in this analysis is the application of the $G$-parity rule \cite{Klempt:2002ap}, which establishes a definite correspondence between the effective interactions of the $K^{(*)} \bar {D}_1/K^{(*)} \bar {D}_2^*$ systems and those of the $K^{(*)} {D}_1/K^{(*)} {D}_2^*$ systems. Specifically, the contributions from the exchange of the particles $\sigma$, $\eta$, and $\rho$ remain identical in both types of systems. In contrast, the contributions arising from the $\pi$ and $\omega$ exchanges differ in sign between two cases \cite{Klempt:2002ap}. Therefore, building on the OBE effective interactions previously derived for the $K^{(*)} \bar {D}_1/K^{(*)} \bar {D}_2^*$ systems, we can apply the $G$-parity rule to obtain the corresponding OBE effective interactions for the $K^{(*)}{D}_1/K^{(*)}{D}_2^*$ systems. With these effective interactions, we proceed to solve the coupled-channel Schr$\ddot{\rm o}$dinger equation incorporating both the $S$-$D$ wave mixing and coupled-channel effects to investigate the $T_{c \bar s}$-type charmed-strange molecular tetraquark candidates composed by the $K^{(*)}$ and $D_1/D_2^*$ mesons.

For the $K D_1$ system with $I(J^P)=0(1^-)$, it involves the effective interactions mediated by the particles $\sigma$, $\omega$, and $\rho$, both of which provide the attractive forces \cite{Chen:2017vai}. As a result, the total effective interaction in the $K D_1$ system with $I(J^P)=0(1^-)$ is stronger than that in the $K \bar D_1$ system with $I(J^P)=0(1^-)$. Nevertheless, in both the single-channel analyses and the analyses incorporating the $S$-$D$ wave mixing effects, no loosely bound state solution is found when the cutoff parameter varies within the range of 0.8-2.0 GeV. 

Similar to the binding behavior observed in the $K D_1$ system with $I(J^P)=0(1^-)$, the $K D_2^*$ system with $I(J^P)=0(2^-)$ also fails to support a loosely bound state under both the single-channel and $S$-$D$ wave mixing scenarios with the cutoff parameter $\Lambda$ varying between 0.8 and 2.0 GeV. 

\begin{figure*}[htbp]
  \centering
  \includegraphics[width=\textwidth]{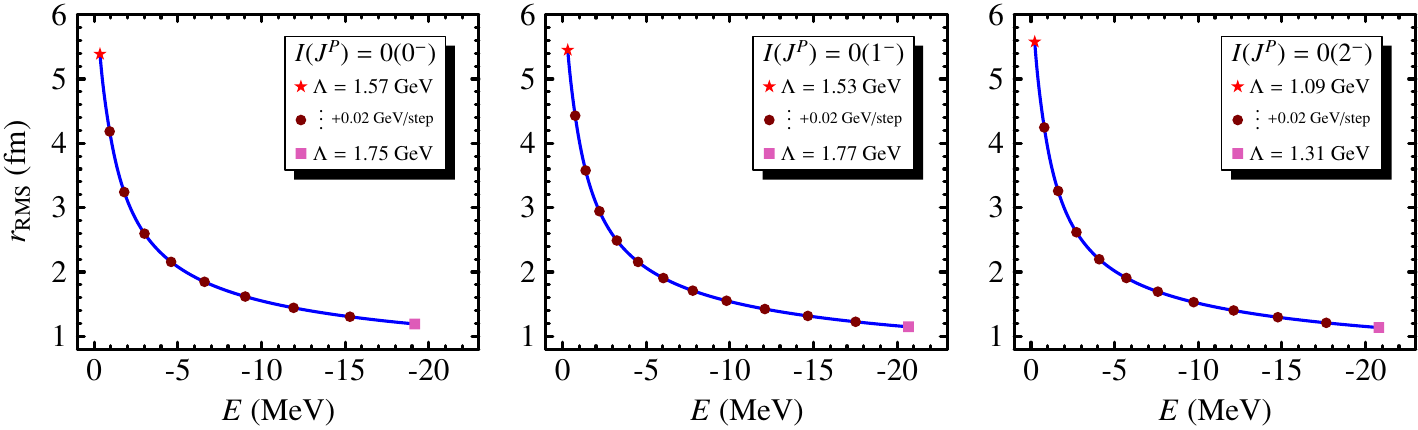}
  \caption{Loosely bound states obtained for the $K^* D_1$ system with the cutoff parameter $\Lambda$ ranging from 0.8 to 2.0 GeV, taking into account the $S$-$D$-wave mixing effects. The star in the upper-left region denotes the loosely bound state that appears at the smallest cutoff value where binding begins, with the corresponding $\Lambda$ value explicitly given. The points from upper-left to lower-right indicates that in each step of the calculation, the cutoff $\Lambda $ increases by 0.02 GeV. The box in the lower-right region indicates the loosely bound solution with a binding energy of about 20 MeV and a root-mean-square radius near 1 fm, also accompanied by its corresponding $\Lambda$ value.}\label{KstD1}
\end{figure*}

In Fig.~\ref{KstD1}, we summarize the obtained loosely bound state solutions for the $K^* D_1$ system by varying the cutoff parameter between 0.8 and 2.0 GeV with the $S$-$D$ wave mixing effects included. The star in the upper-left region denotes the loosely bound state that appears at the smallest cutoff value where binding begins, with the corresponding $\Lambda$ value explicitly given. The box in the lower-right region indicates the loosely bound solution with a binding energy of about 20 MeV and a root-mean-square radius near 1 fm, also accompanied by its corresponding $\Lambda$ value. We also mark the evolution of the binding properties as the cutoff parameter is progressively increased. The points from upper-left to lower-right indicates that in each step of the calculation, the cutoff $\Lambda $ increases by 0.02 GeV. From the Fig.~\ref{KstD1}, it is evident that both the binding energy and the RMS radius exhibit a monotonic dependence on the cutoff parameter. As the cutoff parameter increases, the discussed system becomes more deeply bound, and the RMS radius decreases accordingly.

As illustrated in Fig.~\ref{KstD1}, our analysis reveals that the $K^* D_1$ systems with $I(J^P)=0(0^-,\,1^-,\,2^-)$ can emerge the loosely bound state solutions when the cutoff parameters are set to 1.57, 1.53, and 1.09 GeV, respectively. Furthermore, when the cutoff parameters are increased to 1.76, 1.77, and 1.31 GeV, the $K^* D_1$ systems with $I(J^P)=0(0^-,\,1^-,\,2^-)$ exhibit the binding energies of approximately 20 MeV and the RMS radii of about 1 fm. Based on the systematic numerical study presented above, we conclude that the $K^* D_1$ systems with $I(J^P)=0(0^-,\,1^-,\,2^-)$ are able to exist the loosely bound state solutions for reasonably chosen values of the cutoff parameter. 

\begin{figure*}[htbp]
  \centering
  \includegraphics[width=\textwidth]{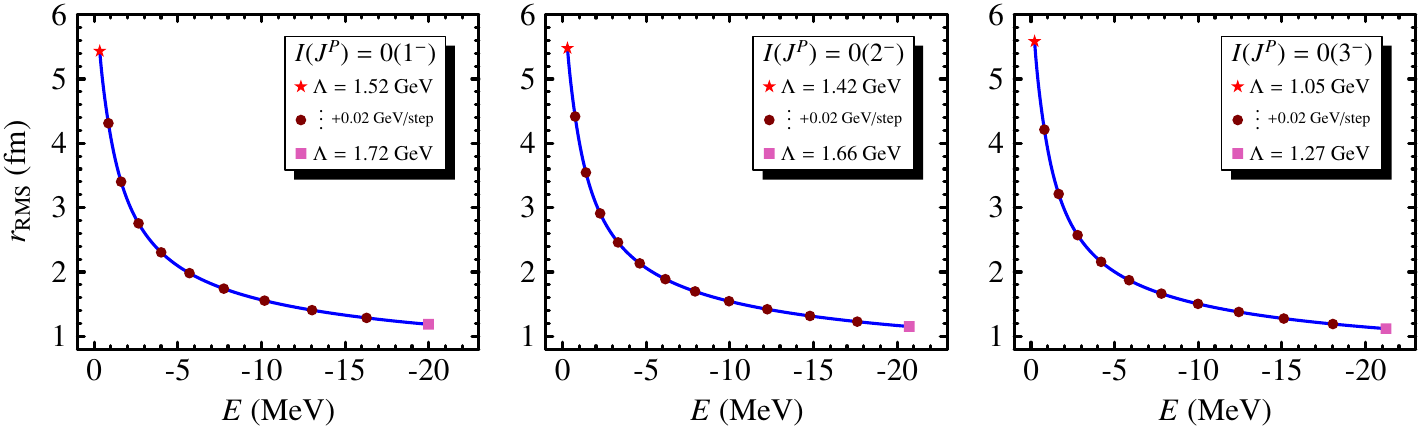}
  \caption{The obtained loosely bound states for the $K^* D_2^*$ system with the cutoff parameter $\Lambda$ varying from 0.8 to 2.0 GeV, including the $S$-$D$-wave mixing effects. The star in the upper-left region marks the loosely bound state emerging at the smallest cutoff value where binding occurs, with the corresponding $\Lambda$ explicitly indicated. The points from upper-left to lower-right indicates that in each step of the calculation, the cutoff $\Lambda $ increases by 0.02 GeV. The box in the lower-right region indicates the loosely bound solution with a binding energy of about 20 MeV and a root-mean-square radius near 1 fm, also accompanied by its corresponding $\Lambda$ value.}\label{KstD2}
\end{figure*}

In a manner analogous to the loosely bound state solutions for the $K^* D_1$ system presented in Fig.~\ref{KstD1}, we display in Fig.~\ref{KstD2} the corresponding loosely bound state solutions for the $K^* D_2^*$ system, obtained by varying the cutoff parameter within the range of 0.8 to 2.0 GeV with the $S$-$D$ wave mixing effects included.

As shown in Fig.~\ref{KstD2}, the $K^* D_2^*$ systems with $I(J^P)=0(1^-,\,2^-,\,3^-)$ support the existence of the loosely bound state solutions once the cutoff parameters reach $\Lambda=$1.52, 1.42, and 1.05 GeV, respectively. At these cutoff parameters, the loosely bound state solutions just appear, characterized by the binding energies of only on the order of tenths of an MeV and the corresponding RMS radii of approximately 5 fm. As the cutoff parameter increases, the binding energies of these bound states increases monotonically, while their RMS radii decrease. This monotonic behavior is consistent with the trend observed in the $K^* D_1$ system. When the cutoff parameters are increased to $\Lambda=$1.73, 1.66, and 1.27 GeV for the three systems, respectively, the binding energies reach approximately 20 MeV and the RMS radii shrink to about 1 fm.

\renewcommand\tabcolsep{0.33cm}
\renewcommand{\arraystretch}{1.50}
\begin{table}[!htbp]
\caption{The obtained loosely bound state solutions for the coupled $K D_1/K^* D_1/K^* D_2^*$ system with $I(J^P)=0(1^-)$. The $P$ are probabilities of various components (in units of \%).}\label{KD1}
\begin{tabular*}{86mm}{@{\extracolsep{\fill}}cccc}\toprule[1pt]\toprule[1pt]
$\Lambda~(\rm{GeV})$ &$E~(\rm {MeV})$ &$r_{\rm RMS}~(\rm {fm})$& $P(K D_1/K^* D_1/K^* D_2^*)$\\
\midrule[0.75pt]
1.01&$-0.83$ &4.64&\textbf{96.60}/3.32/0.08\\
1.04&$-9.52$ &1.66&\textbf{89.37}/10.39/0.24\\
1.06&$-20.34$ &1.16&\textbf{84.94}/14.73/0.33\\
\bottomrule[1pt]\bottomrule[1pt]
\end{tabular*}
\end{table}

For the coupled $K D_1/K^* D_1/K^* D_2^*$ system with $I(J^P)=0(1^-)$, a loosely bound state solution emerges, as listed in Table~\ref{KD1}. At a cutoff value of 1.01 GeV, the binding energy and the RMS radius of the coupled $K D_1/K^* D_1/K^* D_2^*$ system with $I(J^P)=0(1^-)$ are determined to be $0.83$ MeV and 4.64 fm, respectively, with the $K D_1$ channel playing the dominant role. Increasing the cutoff to 1.06 GeV, the binding energy of the coupled $K D_1/K^* D_1/K^* D_2^*$ system with $I(J^P)=0(1^-)$ reaches approximately 20.34 MeV and the RMS radius is about 1.16 fm. In this scenario, although the $K D_1$ channel remains dominant, the $K^* D_1$ channel also contributes significantly, accounting for over 14\% of the system's composition.

\renewcommand\tabcolsep{0.33cm}
\renewcommand{\arraystretch}{1.50}
\begin{table}[!htbp]
\caption{The obtained loosely bound state solutions for the coupled $K D_2^*/K^* D_1/K^* D_2^*$ system with $I(J^P)=0(2^-)$. The $P$ are probabilities of various components (in units of \%).}\label{KD2}
\begin{tabular*}{86mm}{@{\extracolsep{\fill}}cccc}\toprule[1pt]\toprule[1pt]
$\Lambda~(\rm{GeV})$ &$E(\rm {MeV})$ &$r_{\rm RMS}~(\rm {fm})$& $P(K D_2^*/K^* D_1/K^* D_2^*)$\\
\midrule[0.75pt]
1.41&$-1.03$ &4.37&\textbf{97.51}/0.02/2.47\\
1.45&$-8.77$ &1.72&\textbf{92.24}/0.03/7.73\\
1.48&$-20.10$ &1.15&\textbf{87.78}/0.03/12.19\\
\bottomrule[1pt]\bottomrule[1pt]
\end{tabular*}
\end{table}

As summarized in Table~\ref{KD2}, when the coupled-channel effects are incorporated, a loosely bound state solution emerges for the coupled $K D_2^*/K^* D_1/K^* D_2^*$ system with $I(J^P)=0(2^-)$. At a cutoff parameter of $\Lambda= 1.41\,{\rm GeV}$, the coupled $K D_2^*/K^* D_1/K^* D_2^*$ system with $I(J^P)=0(2^-)$ can form a loosely bound state with the binding energy and the RMS radius of 1.41 MeV and 4.37 fm, respectively. Furthermore, when the binding energy of the coupled $K D_2^*/K^* D_1/K^* D_2^*$ system with $I(J^P)=0(2^-)$ is less than 20 MeV, the $K D_1$ channel dominates this bound state, contributing more than 87\%, followed by the $K^* D_2^*$ channel as the second most significant component. 

Following a similar line of analysis as for the $T_{\bar{c}\bar{s}}$-type charmed-strange molecular tetraquark candidates composed of $K^{(*)}$ and $\bar{D}_1/\bar{D}_2^*$ mesons, we suggest that the $K^* D_1$ states with $I(J^P)=0(0^-,\,1^-,\,2^-)$ and the $K^* D_2^*$ states with $I(J^P)=0(1^-,\,2^-,\,3^-)$ can be regarded as the most promising candidates of the $T_{c \bar s}$-type charmed-strange molecular tetraquarks. In contrast, the coupled $K D_1/K^* D_1/K^* D_2^*$ system with $I(J^P)=0(1^-)$ and the coupled $K D_2^*/K^* D_1/K^* D_2^*$ system with $I(J^P)=0(2^-)$ can only be considered as the possible candidates of the $T_{c \bar s}$-type charmed-strange molecular states. 

\renewcommand\tabcolsep{0.27cm}
\renewcommand{\arraystretch}{1.50}
\begin{table}[!htbp]
\centering
\caption{The obtained loosely bound state solutions for the $I(J^P)=0(0^-)$ $K^* D_1$ state and the $I(J^P)=0(1^-)$ $K^* D_1$ state when a common cutoff parameter is adopted.}\label{samecutoff}
\begin{tabular}{c|cc|cc}\toprule[1.0pt]\toprule[1.0pt]
\multirow{2}{*}{$\Lambda~(\rm{GeV})$}&\multicolumn{2}{c|}{$0(0^-)$ $K^* D_1$ state}&\multicolumn{2}{c}{$0(1^-)$ $K^* D_1$ state}\\
&$E~(\rm {MeV})$ &$r_{\rm RMS}~(\rm {fm})$&$E~(\rm {MeV})$ &$r_{\rm RMS}~(\rm {fm})$\\
\midrule[0.75pt]
1.57&$-0.35$&5.39&$-1.38$&3.58\\
1.62&$-2.36$&2.89&$-3.85$&2.31\\
1.67&$-6.60$&1.85&$-7.78$&1.71\\
\bottomrule[1.0pt]\bottomrule[1.0pt]
\end{tabular}
\end{table}

In the above discussion, we scanned the cutoff parameter over a reasonable range to search for the loosely bound states in the charmed-strange tetraquark systems with $K^{(*)}$ and $T$-doublet charmed or anticharmed meson, and consequently predicted a series of possible hadronic molecular candidates. It should be noted that, within the OBE model, different cutoff values were employed in the $I(J^P)=1/2(1/2^-)$ and $1/2(3/2^-)$ $\Sigma_c \bar D^*$ molecular picture to simultaneously reproduce the binding properties of the experimentally observed $P_c (4440)$ and $P_c(4457)$ \cite{Chen:2019asm}. Therefore, in the present analysis, we do not require a common cutoff for all spin-partner states. Nevertheless, it is physically instructive to explore the consequences of imposing the same cutoff parameter on these spin partners. Taking the $I(J^P)=0(0^-)$ $K^* D_1$ state and the $I(J^P)=0(1^-)$ $K^* D_1$ state as illustrative examples (see Table \ref{samecutoff}), we find that their binding behaviors are different under an identical cutoff, and the $I(J^P)=0(1^-)$ $K^* D_1$ state is bound more deeply than the $I(J^P)=0(0^-)$ $K^* D_1$ state. This difference can be primarily attributed to the spin-dependent interactions, which is reminiscent of the findings reported in Ref. \cite{Brambilla:2024imu}.

\section{Discussion and conclusion}\label{sec4}

In this work, we first investigate the $T_{\bar{c}\bar{s}}$-type charmed-strange molecular tetraquark candidates formed by the $K^{(*)}$ and $\bar{D}_1/\bar{D}_2^*$ mesons. These systems are unambiguously exotic due to their flavor content $\bar{c}\bar{s}qq$. The effective interactions between the constituent mesons are derived within the OBE model, incorporating exchanges of the $\sigma$, $\pi$, $\eta$, $\rho$, and $\omega$ mesons, and accounting for both the $S$-$D$ wave mixing and coupled-channel effects. Our results indicate that the $K^* \bar D_1$ states with $I(J^P)=0(0^-,\,1^-)$ and the $K^* \bar D_2^*$ states with $I(J^P)=0(1^-,\,2^-)$ represent the most promising candidates of the $T_{\bar{c}\bar{s}}$-type charmed-strange molecular states. In contrast, the coupled $K \bar D_1 / K^* \bar D_1 / K^* \bar D_2^*$ system with $I(J^P)=0(1^-)$ and the coupled $K \bar D_2^* / K^* \bar D_1 / K^* \bar D_2^*$ system with $I(J^P)=0(2^-)$ can only be regarded as the possible candidates of the $T_{\bar{c}\bar{s}}$-type charmed-strange molecular tetraquarks. Given the limited resources, we recommend prioritizing experimental searches for the most promising candidates identified above.

Building on these findings in the $K^{(*)} \bar{D}_1 / K^{(*)} \bar{D}_2^*$ systems, we extend our formalism to the $K^{(*)} D_1 / K^{(*)} D_2^*$ systems. The effective interactions in the latter are related to those in the former via the $G$-parity rule, which provides a key theoretical constraint. This approach allows us to make predictions for the $T_{c \bar{s}}$-type charmed-strange molecular candidates composed of $K^{(*)}$ and $D_1/D_2^*$ mesons. Our numerical results suggest that the $K^* D_1$ states with $I(J^P)=0(0^-,\,1^-,\,2^-)$ and the $K^* D_2^*$ states with $I(J^P)=0(1^-,\,2^-,\,3^-)$ can be regarded as the most promising candidates of the $T_{c \bar s}$-type charmed-strange molecular states. In contrast, the coupled $K D_1/K^* D_1/K^* D_2^*$ system with $I(J^P)=0(1^-)$ and the coupled $K D_2^*/K^* D_1/K^* D_2^*$ system with $I(J^P)=0(2^-)$ can only be considered as the possible candidates of the $T_{c \bar s}$-type charmed-strange molecular states. Experimental identification of these predicted charmed-strange molecular states formed by a $K^{(*)}$ meson and a (anti-)charmed meson in the $T$-doublet would provide a critical test of the molecular interpretation for several observed exotic hadrons and substantially enrich the family of the charmed-strange molecular states. We therefore strongly encourage future experimental searches for these states, particularly at facilities such as LHCb and Belle II. Promising production mechanisms include $B$-meson weak decays and high-energy proton-proton collisions.

Although this work focuses on the charmed-strange molecular states composed of $S$-wave kaons and $P$-wave (anti-)charmed mesons in the $T$-doublet, other configurations-such as compact multiquark states-differ fundamentally from the molecular picture. A critical question is how to distinguish the configurations between these two types of hadrons.

\begin{figure}[htbp]
\centering
\includegraphics[width=4.7cm]{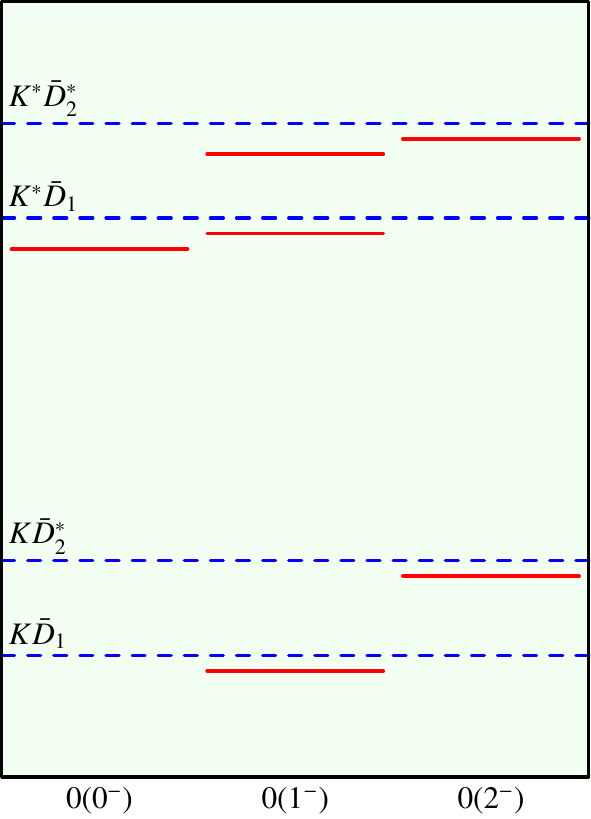}\\
\caption{Characteristic spectra of the $T_{\bar c \bar s}$-type molecular candidates from the $K^{(*)}\bar{D}_1/K^{(*)}\bar{D}_2^*$ interactions.}\label{Characteristicspectra}
\end{figure}

A key observation from the identification of the observed hidden-charm $P_c$ states \cite{Aaij:2019vzc} as the molecular-type exotic states is that their mass spectrum closely matches the predicted characteristic mass spectrum of the hidden-charm molecular-type pentaquark systems \cite{Li:2014gra,Karliner:2015ina,Wu:2010jy,Wang:2011rga,Yang:2011wz,Wu:2012md,Chen:2015loa}. As shown in Fig.~\ref{Characteristicspectra}, there exists a characteristic mass spectrum for the charmed-strange molecular states composed of $S$-wave kaons and $P$-wave anticharmed mesons in the $T$-doublet, which can serve as a valuable criterion for distinguishing different possible hadronic configurations.

In addition, as demonstrated in prior studies \cite{Kalashnikova:2009gt,Baru:2010ww,Chen:2021tad,Chen:2022ddj}, the line-shape analysis has proven to be an effective approach for discriminating between different hadronic configurations. Further investigations along this direction will also be valuable in future work. It is worth noting that the line-shape analyses of the $X_1(2900)$ as a $\bar{D}_1 K$ molecule have already been discussed in Refs. \cite{Chen:2021tad,Chen:2022ddj}.

\section*{Acknowledgement}

This work is supported by the Natural Science Foundation of Gansu Province (No. 26RCKA012 and No. 25JRRA799), the National Natural Science Foundation of China under Grants No. 12335001, No. 12405097, No. 12405098, and No. 12247101, the ‘111 Center’ under Grant No. B20063, the fundamental Research Funds for the Central Universities (lzujbky-2023-stlt01), Lanzhou City High-Level Talent Funding, and the Talent Scientific Fund of Lanzhou University.

\appendix

\section{OBE effective interactions in the coordinate space for the $K^{(*)}\bar T$ systems}\label{app01}

This Appendix summarizes the effective interactions obtained via the OBE model for the $K^{(*)}\bar T$ systems. Following the standard formalism outlined in Sec. \ref{sec2} for deriving the OBE effective interactions in the coordinate space, we present the explicit expressions for the OBE effective interactions in the coordinate space for the $K^{(*)}\bar T$ systems in the following:
\begin{eqnarray*}
{V}^{K \bar {D}_1\to K \bar {D}_1}&=&g_{\sigma}g^{\prime\prime}_{\sigma}\mathcal{O}_1[J]Y_{\sigma}-\frac{1}{2}\beta\beta^{\prime\prime}g^2_V\mathcal{O}_1[J]\mathcal{G}_IY_V,\\
{V}^{K \bar {D}_2^*\to K \bar {D}_2^*}&=&g_{\sigma}g^{\prime\prime}_{\sigma}\mathcal{O}_2[J]Y_{\sigma}-\frac{1}{2}\beta\beta^{\prime\prime}g^2_V\mathcal{O}_2[J]\mathcal{G}_IY_V,\\
{V}^{K^* \bar {D}_1\to K^* \bar {D}_1}&=&g_{\sigma}g^{\prime\prime}_{\sigma}\mathcal{O}_{3}[J]Y_{\sigma}\nonumber\\
    &&+\frac{5gk}{18f^2_{\pi}}\left(\mathcal{O}_{4}[J]\mathcal{Z}_r+\mathcal{O}_{5}[J]\mathcal{T}_r\right)\mathcal{H}_IY_P\nonumber\\
    &&-\frac{1}{2}\beta\beta^{\prime\prime}g^2_V\mathcal{O}_{3}[J]\mathcal{G}_IY_V\nonumber\\
    &&+\frac{5}{9}\lambda\lambda^{\prime\prime}g^2_V\left(2\mathcal{O}_{4}[J]\mathcal{Z}_r-\mathcal{O}_{5}[J]\mathcal{T}_r\right)\mathcal{G}_IY_V,\\
{V}^{K^* \bar {D}_2^*\to K^* \bar {D}_2^*}&=&g_{\sigma}g^{\prime\prime}_{\sigma}\mathcal{O}_{6}[J]Y_{\sigma}\nonumber\\
    &&+\frac{gk}{3f^2_{\pi}}\left(\mathcal{O}_{7}[J]\mathcal{Z}_r+\mathcal{O}_{8}[J]\mathcal{T}_r\right)\mathcal{H}_IY_P\nonumber\\
    &&-\frac{1}{2}\beta\beta^{\prime\prime}g^2_V\mathcal{O}_{6}[J]\mathcal{G}_IY_V\nonumber\\
    &&+\frac{2}{3}\lambda\lambda^{\prime\prime}g^2_V\left(2\mathcal{O}_{7}[J]\mathcal{Z}_r-\mathcal{O}_{8}[J]\mathcal{T}_r\right)\mathcal{G}_IY_V,\\
{V}^{K \bar {D}_1\to K^* \bar {D}_1}&=&\frac{5gk}{6f^2_{\pi}}\left(\mathcal{O}_{9}[J]\mathcal{Z}_r+\mathcal{O}_{10}[J]\mathcal{T}_r\right)\mathcal{H}_IY_{P}^{1}\nonumber\\
    &&-\frac{5}{9}\lambda\lambda^{\prime\prime}g^2_V\left(2\mathcal{O}_{9}[J]\mathcal{Z}_r-\mathcal{O}_{10}[J]\mathcal{T}_r\right)\mathcal{G}_IY_{V}^{1},\\
{V}^{K \bar {D}_1\to K^* \bar {D}_2^*}&=&-\frac{gk}{3\sqrt{6}f^2_{\pi}}\left(\mathcal{O}_{11}[J]\mathcal{Z}_r+\mathcal{O}_{12}[J]\mathcal{T}_r\right)\mathcal{H}_IY_{P}^{2}\nonumber\\
    &&-\frac{\sqrt{2}}{3\sqrt{3}}\lambda\lambda^{\prime\prime}g^2_V\left(2\mathcal{O}_{11}[J]\mathcal{Z}_r-\mathcal{O}_{12}[J]\mathcal{T}_r\right)\mathcal{G}_IY_{V}^{2},\\
{V}^{K \bar {D}_2^*\to K^*\bar {D}_1}&=&-\frac{gk}{3\sqrt{6}f^2_{\pi}}\left(\mathcal{O}_{13}[J]\mathcal{Z}_r+\mathcal{O}_{14}[J]\mathcal{T}_r\right)\mathcal{H}_IY_{P}^{3}\nonumber\\
    &&-\frac{\sqrt{2}}{3\sqrt{3}}\lambda\lambda^{\prime\prime}g^2_V\left(2\mathcal{O}_{13}[J]\mathcal{Z}_r-\mathcal{O}_{14}[J]\mathcal{T}_r\right)\mathcal{G}_IY_{V}^{3},\\
{V}^{K \bar {D}_2^*\to K^* \bar {D}_2^*}&=&\frac{gk}{3f^2_{\pi}}\left(\mathcal{O}_{15}[J]\mathcal{Z}_r+\mathcal{O}_{16}[J]\mathcal{T}_r\right)\mathcal{H}_IY_{P}^{4}\nonumber\\
    &&-\frac{2}{3}\lambda\lambda^{\prime\prime}g^2_V\left(2\mathcal{O}_{15}[J]\mathcal{Z}_r-\mathcal{O}_{16}[J]\mathcal{T}_r\right)\mathcal{G}_IY_{V}^{4},\\
{V}^{K^*\bar {D}_1\to K^* \bar {D}_2^*}&=&\frac{gk}{3\sqrt{6}f^2_{\pi}}\left(\mathcal{O}_{17}[J]\mathcal{Z}_r+\mathcal{O}_{18}[J]\mathcal{T}_r\right)\mathcal{H}_IY_{P}^{5}\nonumber\\
    &&+\frac{\sqrt{2}}{3\sqrt{3}}\lambda\lambda^{\prime\prime}g^2_V\left(2\mathcal{O}_{17}[J]\mathcal{Z}_r-\mathcal{O}_{18}[J]\mathcal{T}_r\right)\mathcal{G}_IY_{V}^{5},
\end{eqnarray*}
where the function $Y_{\mathcal{E}}^i$ is defined as
\begin{eqnarray*}
Y_{\mathcal{E}}^i= \frac{e^{-m_{\mathcal{E}}^i r}-e^{-\Lambda^ir}}{4\pi r}-\frac{{\Lambda^i}^2-{m_{\mathcal{E}}^i}^2}{8\pi\Lambda^i}e^{-\Lambda^ir},
\end{eqnarray*}
with $m_{\mathcal{E}}^i=\sqrt{m_{\mathcal{E}}^2-{q_0^i}^2}$ and $\Lambda^i=\sqrt{\Lambda^2-{q_0^i}^2}$ $(i=1,...,5)$. The zeroth components of
exchange meson momentum $q_0^i$ are derived in Ref. \cite{Wang:2023ael} and take the values $q_0^1=0.084\,{\rm GeV}$, $q_0^2=0.055\,{\rm GeV}$, $q_0^3=0.113\,{\rm GeV}$, $q_0^4=0.083\,{\rm GeV}$, and $q_0^5=0.028\,{\rm GeV}$. We note that the exchanged pion remains off-shell in all processes considered in the present work. The operators $\mathcal{Z}_r$ and $\mathcal{T}_r$ acting on $Y_{\mathcal{E}}^i$ are given by $\mathcal{Z}_r = \frac{1}{r^2}\frac{\partial}{\partial r}r^2\frac{\partial}{\partial r}$ and $\mathcal{T}_r = r\frac{\partial}{\partial r}\frac{1}{r}\frac{\partial}{\partial r}$. The $K^{(*)}\bar T$ systems can have the isospin quantum numbers $I=0$ or $1$, and the isospin factors in the corresponding OBE effective interactions are defined as:
\begin{eqnarray*}
\mathcal{H}_0Y_P=-\dfrac{3}{2}Y_{\pi}+\dfrac{1}{6}Y_{\eta},~~~~~~~\mathcal{H}_1Y_P=\dfrac{1}{2}Y_{\pi}+\dfrac{1}{6}Y_{\eta},\\
\mathcal{G}_0Y_V=-\dfrac{3}{2}Y_{\rho}+\dfrac{1}{2}Y_{\omega},~~~~~~~\mathcal{G}_1Y_V=\dfrac{1}{2}Y_{\rho}+\dfrac{1}{2}Y_{\omega}.
\end{eqnarray*}
Here, $\mathcal{H}_{I}$ and $\mathcal{G}_{I}$ denote the isospin factors for the $K^{(*)}\bar T$ systems, whose specific values are determined from the corresponding flavor wave functions constructed earlier. 

Furthermore, we provide the definitions of the operators $\mathcal{O}_{i}[J]$ employed in this work:
\begin{eqnarray*}
\mathcal{O}_{1}[J]&=&{\bm\epsilon^{\dagger}_4}\cdot{\bm\epsilon_2},\\
\mathcal{O}_{2}[J]&=&\sum_{m,n}^{a,b}\mathbb{C}^{2,m+n}_{1,m;1,n}\mathbb{C}^{2,a+b}_{1,a;1,b}\left({\bm\epsilon^{\dagger}_{4m}}\cdot{\bm\epsilon_{2a}}\right)\left({\bm\epsilon^{\dagger}_{4n}}\cdot {\bm\epsilon_{2b}}\right),\\
\mathcal{O}_{3}[J]&=&\left({\bm\epsilon^{\dagger}_3}\cdot{\bm\epsilon_1}\right)\left({\bm\epsilon^{\dagger}_4}\cdot{\bm\epsilon_2}\right),\\
\mathcal{O}_{4}[J]&=&\left({\bm\epsilon^{\dagger}_3}\times{\bm\epsilon_1}\right)\cdot\left({\bm\epsilon^{\dagger}_4}\times{\bm\epsilon_2}\right),\\
\mathcal{O}_{5}[J]&=&S({\bm\epsilon^{\dagger}_3}\times{\bm\epsilon_1},{\bm\epsilon^{\dagger}_4}\times{\bm\epsilon_2},\hat{\bm{r}}),\\
\mathcal{O}_{6}[J]&=&\sum_{m,n}^{a,b}\mathbb{C}^{2,m+n}_{1,m;1,n}\mathbb{C}^{2,a+b}_{1,a;1,b}\left({\bm\epsilon^{\dagger}_3}\cdot{\bm\epsilon_1}\right)\left({\bm\epsilon^{\dagger}_{4m}}\cdot{\bm\epsilon_{2a}}\right)\left({\bm\epsilon^{\dagger}_{4n}}\cdot {\bm\epsilon_{2b}}\right),\\
\mathcal{O}_{7}[J]&=&\sum_{m,n}^{a,b}\mathbb{C}^{2,m+n}_{1,m;1,n}\mathbb{C}^{2,a+b}_{1,a;1,b}\left({\bm\epsilon^{\dagger}_{4m}}\cdot{\bm\epsilon_{2a}}\right)\left({\bm\epsilon^{\dagger}_{3}}\times{\bm\epsilon_{1}}\right)\cdot\left({\bm\epsilon^{\dagger}_{4n}}\times {\bm\epsilon_{2b}}\right),\\
\mathcal{O}_{8}[J]&=&\sum_{m,n}^{a,b}\mathbb{C}^{2,m+n}_{1,m;1,n}\mathbb{C}^{2,a+b}_{1,a;1,b}\left({\bm\epsilon^{\dagger}_{4m}}\cdot{\bm\epsilon_{2a}}\right)S({\bm\epsilon^{\dagger}_{3}}\times{\bm\epsilon_{1}},{\bm\epsilon^{\dagger}_{4n}}\times {\bm\epsilon_{2b}},\hat{\bm{r}}),\\
\mathcal{O}_{9}[J]&=&i{\bm\epsilon^{\dagger}_3}\cdot\left({\bm\epsilon^{\dagger}_4}\times{\bm\epsilon_2}\right),\\
\mathcal{O}_{10}[J]&=&S({\bm\epsilon^{\dagger}_3},i{\bm\epsilon^{\dagger}_4}\times{\bm\epsilon_2},\hat{\bm{r}}),\\
\mathcal{O}_{11}[J]&=&\sum_{m,n}\mathbb{C}^{2,m+n}_{1,m;1,n}\left({\bm\epsilon^{\dagger}_{4m}}\cdot{\bm\epsilon_{2}}\right)\left({\bm\epsilon^{\dagger}_{4n}}\cdot {\bm\epsilon^{\dagger}_{3}}\right),\\
\mathcal{O}_{12}[J]&=&\sum_{m,n}\mathbb{C}^{2,m+n}_{1,m;1,n}\left({\bm\epsilon^{\dagger}_{4m}}\cdot{\bm\epsilon_{2}}\right)S({\bm\epsilon^{\dagger}_{4n}},{\bm\epsilon^{\dagger}_{3}},\hat{\bm{r}}),\\
\mathcal{O}_{13}[J]&=&\sum_{m,n}\mathbb{C}^{2,m+n}_{1,m;1,n}\left({\bm\epsilon^{\dagger}_{4}}\cdot{\bm\epsilon_{2m}}\right)\left({\bm\epsilon^{\dagger}_{3}}\cdot {\bm\epsilon_{2n}}\right),\\
\mathcal{O}_{14}[J]&=&\sum_{m,n}\mathbb{C}^{2,m+n}_{1,m;1,n}\left({\bm\epsilon^{\dagger}_{4}}\cdot{\bm\epsilon_{2m}}\right)S({\bm\epsilon^{\dagger}_{3}},{\bm\epsilon_{2n}},\hat{\bm{r}}),\\
\mathcal{O}_{15}[J]&=&\sum_{m,n}^{a,b}\mathbb{C}^{2,m+n}_{1,m;1,n}\mathbb{C}^{2,a+b}_{1,a;1,b}\left({\bm\epsilon^{\dagger}_{4m}}\cdot{\bm\epsilon_{2a}}\right)\left[i{\bm\epsilon^{\dagger}_{3}}\cdot\left({\bm\epsilon^{\dagger}_{4n}}\times {\bm\epsilon_{2b}}\right)\right],\\
\mathcal{O}_{16}[J]&=&\sum_{m,n}^{a,b}\mathbb{C}^{2,m+n}_{1,m;1,n}\mathbb{C}^{2,a+b}_{1,a;1,b}\left({\bm\epsilon^{\dagger}_{4m}}\cdot{\bm\epsilon_{2a}}\right)S({\bm\epsilon^{\dagger}_{3}},i{\bm\epsilon^{\dagger}_{4n}}\times {\bm\epsilon_{2b}},\hat{\bm{r}}),\\
\mathcal{O}_{17}[J]&=&\sum_{m,n}^{a,b}\mathbb{C}^{2,m+n}_{1,m;1,n}\mathbb{C}^{2,a+b}_{1,a;1,b}\left({\bm\epsilon^{\dagger}_{4m}}\cdot{\bm\epsilon_{2}}\right)\left[i{\bm\epsilon^{\dagger}_{4n}}\cdot\left({\bm\epsilon^{\dagger}_{3}}\times {\bm\epsilon_{1}}\right)\right],\\
\mathcal{O}_{18}[J]&=&\sum_{m,n}^{a,b}\mathbb{C}^{2,m+n}_{1,m;1,n}\left({\bm\epsilon^{\dagger}_{4m}}\cdot{\bm\epsilon_{2}}\right)S({\bm\epsilon^{\dagger}_{4n}},i{\bm\epsilon^{\dagger}_{3}}\times {\bm\epsilon_{1}},\hat{\bm{r}}).
\end{eqnarray*}
The tensor force operator appearing above is defined as $S({\bm x},{\bm y},\hat{\bm{r}})= 3\left(\hat{\bm r} \cdot {\bm x}\right)\left(\hat{\bm r} \cdot {\bm y}\right)-{\bm x} \cdot {\bm y}$. In the practical calculations, the matrix elements of the operators $\langle f|\mathcal{O}_i[J]|i\rangle$ are evaluated by sandwiching them $\mathcal{O}_{i}[J]$ between the spin-orbital wave functions of the initial states $|i\rangle$ and the final states $|f\rangle$. The resulting operator matrix elements $\langle f|\mathcal{O}_i[J]|i\rangle$  for the OBE effective potentials of the $K^{(*)}\bar T$ systems in the coordinate space are listed in Table~\ref{operators}. When solving the coupled-channel Schr$\ddot{\rm o}$dinger equations, the operators appearing in the OBE effective interactions are replaced by their corresponding matrix elements provided in Table~\ref{operators}.
\renewcommand\tabcolsep{0.56cm}
\renewcommand{\arraystretch}{2.0}
\begin{table*}[!htbp]
\caption{The resulting operator matrix elements $\langle f|\mathcal{O}_i[J]|i\rangle$ of the OBE effective interactions in the coordinate space for the $K^{(*)}\bar T$ systems.}\label{operators}
\begin{tabular}{c}\toprule[1pt]\toprule[1pt]
\multicolumn{1}{l}{$\begin{array}{cc} \langle f|\mathcal{O}_1[1]|i\rangle=\rm {diag}(1,1) \\ \langle f|\mathcal{O}_2[2]|i\rangle=\rm {diag}(1,1)\end{array}$~~~~~~~~
$\begin{array}{cc} \langle f|\mathcal{O}_3[0]|i\rangle=\rm {diag}(1,1) \\ \langle f|\mathcal{O}_4[0]|i\rangle=\rm {diag}(2,-1)\end{array}$~~~~~~~~
$\langle f|\mathcal{O}_{5}[0]|i\rangle=\left(\begin{array}{cc} 0 & \sqrt{2} \\ \sqrt{2} & 2\end{array}\right)$~~~~~~~~
$\begin{array}{cc} \langle f|\mathcal{O}_3[1]|i\rangle=\rm {diag}(1,1,1) \\ \langle f|\mathcal{O}_4[1]|i\rangle=\rm {diag}(1,1,-1)\end{array}$}\\
\multicolumn{1}{l}{$\langle f|\mathcal{O}_{5}[1]|i\rangle=\left(\begin{array}{ccc} 0 & -\sqrt{2} &0 \\ -\sqrt{2} & 1 &0 \\ 0&0&1\end{array}\right)$~~~~~~~~
$\begin{array}{cc} \langle f|\mathcal{O}_3[2]|i\rangle=\rm {diag}(1,1,1,1) \\ \langle f|\mathcal{O}_4[2]|i\rangle=\rm {diag}(-1,2,1,-1)\end{array}$~~~~~~~~
$\langle f|\mathcal{O}_{5}[2]|i\rangle=\left(\begin{array}{cccc} 0 & \frac{\sqrt{2}}{\sqrt{5}} & 0 &-\frac{\sqrt{14}}{\sqrt{5}} \\ \frac{\sqrt{2}}{\sqrt{5}} & 0 & 0 &-\frac{2}{\sqrt{7}} \\ 0 & 0 &-1 &0 \\ -\frac{\sqrt{14}}{\sqrt{5}}&-\frac{2}{\sqrt{7}}&0&-\frac{3}{7}\end{array}\right)$}\\
\multicolumn{1}{l}{$\begin{array}{cc} \langle f|\mathcal{O}_{6}[1]|i\rangle=\rm {diag}(1,1,1,1) \\ \langle f|\mathcal{O}_{7}[1]|i\rangle=\rm {diag}(\frac{3}{2},\frac{3}{2},\frac{1}{2},-1)\end{array}$~~~~~~~~
$\langle f|\mathcal{O}_{8}[1]|i\rangle=\left(\begin{array}{cccc} 0&\frac{3}{5\sqrt{2}}&\frac{\sqrt{6}}{\sqrt{5}}&\frac{\sqrt{21}}{5\sqrt{2}} \\ \frac{3}{5\sqrt{2}}&-\frac{3}{10}&\frac{\sqrt{3}}{\sqrt{5}}&-\frac{\sqrt{3}}{5\sqrt{7}}\\\frac{\sqrt{6}}{\sqrt{5}}&\frac{\sqrt{3}}{\sqrt{5}}&\frac{1}{2}&\frac{2}{\sqrt{35}}\\\frac{\sqrt{21}}{5\sqrt{2}}
&-\frac{\sqrt{3}}{5\sqrt{7}}&\frac{2}{\sqrt{35}}&\frac{48}{35}\end{array}\right)$~~~~~~~~
$\begin{array}{cc} \langle f|\mathcal{O}_{6}[2]|i\rangle=\rm {diag}(1,1,1,1)\\ \langle f|\mathcal{O}_{7}[2]|i\rangle=\rm {diag}(\frac{1}{2},\frac{3}{2},\frac{1}{2},-1)\end{array}$}\\
\multicolumn{1}{l}{$\langle f|\mathcal{O}_{8}[2]|i\rangle=\left(\begin{array}{cccc} 0&-\frac{3\sqrt{2}}{5}&-\frac{\sqrt{7}}{\sqrt{10}}&\frac{\sqrt{7}}{5} \\ -\frac{3\sqrt{2}}{5}&\frac{3}{10}&\frac{3}{\sqrt{35}}&-\frac{3\sqrt{2}}{5\sqrt{7}}\\-\frac{\sqrt{7}}{\sqrt{10}}&\frac{3}{\sqrt{35}}&-\frac{3}{14}&\frac{4\sqrt{2}}{7\sqrt{5}}\\\frac{\sqrt{7}}{5}
&-\frac{3\sqrt{2}}{5\sqrt{7}}&\frac{4\sqrt{2}}{7\sqrt{5}}&\frac{12}{35}\end{array}\right)$~~~~~~~~
$\begin{array}{cc} \langle f|\mathcal{O}_{6}[3]|i\rangle=\rm {diag}(1,1,1,1) \\ \langle f|\mathcal{O}_{7}[3]|i\rangle=\rm {diag}(-1,\frac{3}{2},\frac{1}{2},-1)\end{array}$~~~~~~~~
$\langle f|\mathcal{O}_{8}[3]|i\rangle=\left(\begin{array}{cccc} 0&\frac{3}{5\sqrt{2}}&-\frac{1}{\sqrt{5}}&-\frac{4\sqrt{3}}{5} \\ \frac{3}{5\sqrt{2}}&-\frac{3}{35}&-\frac{6\sqrt{2}}{7\sqrt{5}}&-\frac{6\sqrt{6}}{35}\\-\frac{1}{\sqrt{5}}&-\frac{6\sqrt{2}}{7\sqrt{5}}&-\frac{4}{7}&\frac{\sqrt{3}}{7\sqrt{5}}\\-\frac{4\sqrt{3}}{5}
&-\frac{6\sqrt{6}}{35}&\frac{\sqrt{3}}{7\sqrt{5}}&-\frac{22}{35}\end{array}\right)$}\\
\multicolumn{1}{l}{$\langle f|\mathcal{O}_{9}[1]|i\rangle=\sqrt{2}$~~~~~~~~$\langle f|\mathcal{O}_{11}[1]|i\rangle=-\frac{\sqrt{5}}{\sqrt{3}}$~~~~~~~~$\langle f|\mathcal{O}_{17}[1]|i\rangle=-\frac{\sqrt{5}}{\sqrt{6}}$~~~~~~~~$\langle f|\mathcal{O}_{13}[2]|i\rangle=1$~~~~~~~~$\langle f|\mathcal{O}_{15}[2]|i\rangle=\frac{\sqrt{3}}{\sqrt{2}}$~~~~~~~~$\langle f|\mathcal{O}_{17}[2]|i\rangle=-\frac{\sqrt{3}}{\sqrt{2}}$}\\
\bottomrule[1pt]\bottomrule[1pt]
\end{tabular}
\end{table*}

\end{document}